\documentclass[12pt,noshowpacs,nofootinbib,notitlepage,amsmath,superscriptaddress]{revtex4-1}
\usepackage{setspace}
\usepackage{hyperref}
\usepackage{amsfonts}
\usepackage{appendix}
\usepackage{mathtools}
\usepackage{ulem}
\usepackage{ifthen}
\newboolean{PrintVersion}
\setboolean{PrintVersion}{false}
\usepackage{amssymb,amstext} 
\usepackage{graphicx} 
\usepackage[caption=false]{subfig}
\usepackage{wrapfig}
\usepackage{cleveref} 
\usepackage{enumitem}
\usepackage{xcolor}
\usepackage[export]{adjustbox}
\usepackage{float}
\usepackage{booktabs}
\usepackage[section]{placeins}

\ifthenelse{\boolean{PrintVersion}}{   
\hypersetup{	
    citecolor=black,%
    filecolor=black,%
    linkcolor=black,%
    urlcolor=black}
}{} 

\begin{document}

\title{Dyonic Quark Stars in Quasi-Topological Electromagnetism}

\author{Michael Gammon}
\email{mgammon@uwaterloo.ca}
\affiliation{Department of Physics and Astronomy, University of Waterloo, Waterloo, Ontario, Canada, N2L 3G1}

\author{Nicola De Kock}
\email{35288949@mylife.unisa.ac.za}
\affiliation{Department of Mathematical Sciences, University of South Africa, Pretoria, South Africa}

 \author{Robert B. Mann}
\email{rbmann@uwaterloo.ca}
\affiliation{Department of Physics and Astronomy, University of Waterloo, Waterloo, Ontario, Canada, N2L 3G1}

\author{Amos Kubeka}
\email{Kubekas@unisa.ac.za}
\affiliation{Department of Mathematical Sciences, University of South Africa, Pretoria, South Africa}

\begin{abstract}
In this paper we consider quark star solutions to Liu et al.'s \cite{Liu_2019} quasi-topological electromagnetism (QTEM), a recently proposed form of dark energy.  Since the QTEM contribution is trivial for pure electric/magnetic charge, we consider the dyonic case in pure QTEM which does induce (dark) non-trivial dynamics from the non-linear theory. Besides the introduction of a dyonic charge distribution generally pushing the characteristic quark star `hook' shape to larger masses and radii, it also induces a second branch at very large mass and radius for stars with a small dyonic charge ratio. This second set of solutions have a negative pressure envelope  surrounding a positive pressure core. 
As we explore the parameter space these features  interact and evolve in interesting ways, with the two branches eventually merging in $M/R$ space before settling into a characteristic `paperclip' shape as the dyonic charge ratio becomes large.
 
\end{abstract}
\maketitle
\newpage

\section{Introduction}
Recent work by Liu et al. \cite{Liu_2019} introduced the quasi-topological theory of electromagnetism (QTEM), a nonlinear generalization of normal Maxwell electromagnetism which is constructed by considering `basic topological structures' built from the field strength tensor (namely $F \wedge  F$ in four dimensions). Nontrivial dynamics follow from using the squared norms of these objects as the quasi-topological invariants from which the Lagrangian of the theory is built. The full equations of motion end up depending on the sum of a term from normal Maxwell electromagnetism and a non-linear quasi-topological (QT) term, where the latter acts as an effective cosmological constant on the spacetime. Liu et al. suggest that this could provide a possible mechanism for a kind of dark energy to be coupled with other types of matter \cite{Liu_2019}.

Purely  electric or magnetic Reissner-Nordstrom black holes in QTEM are identical to their pure Maxwell counterparts, but the dyonic black hole (carrying both electric and magnetic charge) admits novel solutions with unique properties like multiple photon spheres and up to four real horizons \cite{Liu_2019}. Further work using QTEM has included these dyonic black hole solutions (and their thermodynamic properties) in both Einstein \cite{Cisterna_2020, Barrientos_2022,Li_2022} and Lovelock \cite{ali2025exoticlovelockblackholes, Sekhmani:2022lws,Sekhmani_2023} gravity. QTEM papers have also come out on black string solutions \cite{Cisterna_2020}, geometrically thick tori solutions \cite{zhou2024geometricallythickequilibriumtori}, super extremal black hole solutions \cite{hod2024superextremalblackholesquasitopological}, as well as analyses of chaotic particle motion near QTEM black holes \cite{Lei_2021}.

However, the implications of QTEM for gravitationally bound objects are essentially unexplored. If such dark energy exists, it will necessarily affect gravitational bound states such as stars, neutron stars, and quark stars. 
Here we consider the implications of QTEM for quark stars -- their strong gravitational binding in combination with  the simplicity of their equation of state make them interesting objects of study. 



Recent studies on quark stars more generally have been gaining renewed interest \cite{zhang_2021_unified,zhang_2021_stellar,gammon_2024,gammon_2025_chargedquarkstars4degb,saavedra_2025,banerjee_2021_quark,banerjee_2021_strange}, partially due to a number of astrophysical observations that are in tension with simple neutron star equations of state  \cite{Horvath_2023,Miller_2019,salmi2024,Abbott_2020}, and partially due to new theoretical advances that suggest  non-strange quark matter (QM) could feasibly be the ground state of baryonic matter at  sufficiently high density and temperature \cite{holdom_2018_quark}. In a number of these papers it has been shown that quark matter EOSs can be used to describe astrophysical objects like HESS J1731-347, PSR J0030+0451, PSR J0740+6620, and GW190814 \cite{zhang_2021_unified,zhang_2021_stellar,gammon_2024,gammon_2025_chargedquarkstars4degb}. 



Motivated by the above, we  investigate dyonic quark stars in pure QTEM.  As could be expected from prior work on dyonic black holes, we find that the contributions from the quasi-topological term vanish unless both an electric and magnetic charge are present. As with the black hole case
\cite{Liu_2019}, we also find that the QTEM terms act as an effective cosmological constant on the spacetime, making these stars a source for a type of dark energy. However our most striking result is that non-zero dyonic charge in pure QTEM induces a change in the quark star solution profile,  where a second branch appears at a larger mass and radius; in other words for a range of central pressures we find that there are two values of $R$ satisfying $p(R)=0$, the equation that defines the radius of the star. This change comes from another change in the structure of the pressure function $p(r)$ itself, where after going from $p(r=0)=p_0$ to $p(R)=0$ at some radius $R$ (which we normally treat as the stellar surface), the pressure stays negative briefly before turning upward and crossing 0 again. In this way the spacetime describes a `normal' quark star surrounded by a shell of positive mass density held together by tension (negative pressure), much like a rubber band.
 
The outline of our paper is as follows. In section \ref{sec:theory} we will outline the basic theory underlying quasi-topological electromagnetism, with section \ref{sec:eoms} outlining the relevant equations of motion, \ref{sec:eoscharge} containing the quark matter equation of state (EOS) and charge distribution models, and section \ref{sec:units} discussing units, scaling, and the relevance of the MIT bag constant. Section \ref{sec:results} shows our numerical results in the form of mass/radius and mass/central density diagrams, followed by a discussion on the general features of the parameter space and the differences between the employed charge distribution models. In section \ref{sec:stability} we touch briefly on the stability of these objects before summarizing our results and concluding the paper in section \ref{sec:summary}.

\section{Theory}\label{sec:theory}
\subsection{Quasi-Topological Electromagnetism}\label{sec:qtem}
The standard (charged) Tolman-Oppenheimer-Volkoff (TOV) equations for stellar structure are well-known when derived from Einstein's general theory of relativity (GR) and standard Maxwell electromagnetism \cite{oppenheimer1939,tolman1939,ray2003,ray2004,zhang_2021_stellar}, and the corresponding solutions have been carefully studied as solutions that describe compact stellar bodies. These results are straightforwardly extended to the case of a dyonic (electrically/magnetically charged) quark star in QTEM.

To start, it is useful to define the following irreducible polynomial notations based on the field strength tensor $F$:
\begin{equation}
\begin{aligned} 
F^{(2)} &=  F^\mu_\nu F^\nu_\mu = -F^2, \\ 
F^{(4)} &=  F^\mu_\nu F^\nu_\rho F^\rho_\sigma F^\sigma_\mu, \\
&....\\
F^{(2n)} &= F^{\mu_1}_{\mu_2}F^{\mu_2}_{\mu_3} ... F^{\mu_{2n}}_{\mu_1},
\end{aligned}
\end{equation}
where $F_{\mu\nu} = \nabla_\mu B_\nu - \nabla_\nu B_\mu$, with $B_\mu$ being the dark matter gauge field.
With these definitions, the QTEM   Lagrangian   is given by  \cite{Liu_2019}
\begin{equation}  \mathcal{L}_{EM}= \left( -\alpha_1 F^2 -\alpha_2 ((F^2)^2-2F^{(4)})\right)
\end{equation}
where $\alpha_1$ is the coupling constant of the standard Maxwell term and  $\alpha_2$ that of 
the quasi-topological part. Standard Maxwell electromagnetism is recovered when $\alpha_2=0$, whereas $\alpha_1=0$ gives the equations of motion for pure quasi-topological electromagnetism. The latter is the case we study in this paper, in combination with a matter source term and the Einstein-Hilbert action term:
\begin{equation}
    \mathcal{L}= \mathcal{L}_{EH}+\mathcal{L}_\mathrm{EM}+\mathcal{L}_\mathrm{M} +16\pi J^\mu B_\mu 
\end{equation}
where $\mathcal{L}_{EH} = R/(16 \pi)$ 
$\cal{L}_\mathrm{M}$ is the Lagrangian of a perfect fluid, and $J^\mu$ is the   current density associated with the dark energy.

\subsection{Equations of Motion}\label{sec:eoms}

We begin the derivation of the equations of motion by employing a static, spherically symmetric metric ansatz to characterize the spacetime:
\begin{equation}\label{metanz}
    ds^2 = -e^{2\nu(r)}dt^2 + e^{2\Lambda(r)} dr^2 +r^2 (d\theta^2 +\sin^2\theta\; d\phi^2)
\end{equation}
along with a perfect fluid stress energy tensor
\begin{equation}
    T_{F}^{\mu \nu} = (\rho + P)u^\mu u^\nu - P g^{\mu \nu}.
\end{equation}
In addition, we employ the ansatz 
\begin{equation}
    B= \Phi(r) dt + g(r)\cos\theta d\phi
\end{equation}
for the dark gauge potential $B_\mu$, where $\Phi(r)$ is the usual electric (scalar) potential, $g(r)$ is the magnetic potential, and $\theta$ is the polar angle. The field strength tensor can then be written as
\begin{equation}\label{Ften}
F_{\mu\nu} = \begin{bmatrix}
0 & -\Phi'(r) & 0 & 0\\
\Phi'(r) & 0 & 0 & g'(r)\cos(\theta)\\
0& 0& 0& -g(r)\sin(\theta)\\
0&-g'(r)\cos(\theta)&g(r)\sin(\theta)&0
\end{bmatrix}.
\end{equation}

We notice that in addition to the usual terms, $F_{\mu\nu}$ now contains additional $[r, \phi]$ and $[\theta, \phi]$ components. Now we define the Maxwell/quasi-topological parts of the electromagnetic energy momentum tensor
\begin{equation}
   T^{EM}_{\mu \nu} =\frac{-2}{\sqrt{-g}}\frac{\delta S_{EM}}{\delta g^{\mu\nu}} = \alpha_1 T_{\mu \nu}^{(1)} + \alpha_2 T_{\mu \nu}^{(2)}
\end{equation}
with 
\begin{equation}
\begin{aligned} 
T_{\mu \nu}^{(1)} &=  F_{\mu \rho}F_\nu^{\; \rho}-\frac{1}{4}F^2g_{\mu \nu}, \\
  T_{\mu \nu}^{(2)} &= 2F^2F_{\mu\rho}F_\nu^{\; \rho} - 4 F_{\mu \rho}F^\rho_{\; \sigma} F^\sigma_{\; \lambda} F^\lambda_{\; \nu} - \frac{1}{4}\left((F^2)^2-2 F^{(4)}\right)g_{\mu \nu},
\end{aligned}
\end{equation}
so that 
$T^{(1)}$ is the usual stress-energy tensor  of Maxwell electromagnetism, whereas $T^{(2)}$ is the stress-energy contribution resulting from the pure quasi-topological term. It is easy to check that $T^{(1)}$ is traceless, but $T^{(2)}$ is not.

 The QTEM field equations are 
\begin{equation}\label{genmax}
   \alpha_1\nabla_\mu F^{\mu \nu} +  \alpha_2 \nabla_\mu \bar{F}^{\mu \nu} = 4 \pi J^\nu.
\end{equation}
(where $\bar{F}^{\mu \nu} =2 F^2F^{\mu\nu}-4F^{\mu\rho}F^\sigma_{\; \rho}F_\sigma^{\; \nu}$), in addition to the Bianchi identity  \begin{equation}\label{Bianchi}
\nabla_{[\mu}F_{\nu\rho]} = 0.
\end{equation}

Inserting \eqref{Ften} into
the $t$-component of 
\eqref{genmax} 
yields a differential equation for $\frac{d}{dr}\Phi(r)$. Employing the ansatz
\begin{equation}\label{eq:dscalar}   \frac{d}{dr}\Phi(r) =\frac{q(r)e^{\nu(r)+\Lambda(r)}}{\alpha_1 r^2+\frac{4 \alpha_2 g(r)^2}{r^2}},
\end{equation}
yields for $\alpha_1=0$ the general solution  
\begin{equation}\label{eq:littleQsol}
\begin{aligned}
    \frac{q(r)}{g(r)}&=4\pi  \int J^0 r^2 e^{\Lambda(r)+\nu(r)}\frac{dr}{g(r)}+C \\
    &=4\pi  \int \tilde{\rho}_x(r) r^2 e^{\Lambda(r)} dr + C 
\end{aligned}
\end{equation}
where $\tilde{\rho}_x(r)=\frac{J^0}{g(r)} e^{\nu(r)}$ is the (dark) electric charge density and $C$ is a constant of integration. 
We note that although only the
ratio $ \frac{q(r)}{g(r)}$ is determined from the field equations when $\alpha_1=0$, 
this equation is qualitatively the same as the usual differential equation for the electric charge $q(r)$ for standard charged stars in Maxwell theory (namely, $\frac{dq(r)}{dr}=4 \pi r^2 \rho_e(r)e^{\Lambda(r)}$ where $\rho_e(r)=J^t(r)e^{\nu(r)}$) \cite{zhang2021}.


The next task is to choose a form for the radial metric function $e^{2\Lambda(r)}$. Prior work on quasi-topological electromagnetism \cite{Liu_2019} has shown that the static, spherically symmetric 
electro-vacuum solution for the metric function is:
\begin{equation}\label{vacanz}
    e^{-2\Lambda(r)}=1-\frac{2M}{r}+\frac{\alpha_1 G^2}{r^2}-\frac{Q^2 r^2}{12 G^2 \alpha_2},
\end{equation}
where $M$, $G$, and $Q$ are the total enclosed mass, magnetic charge, and electric charge, respectively. This suggests that we adopt 
\begin{equation}\label{Lamanz}
    e^{-2\Lambda(r)}=1-\frac{2m(r)}{r} +\frac{\alpha_1 g(r)^2}{r^2}-\frac{q(r)^2 r^2}{12 g(r)^2 \alpha_2}
\end{equation}
 for the metric function $\Lambda(r)$.   
With this, on computing the $tt$ and $rr$ components of the field equations, we find that they have a non-vanishing $\theta$ dependence, 
in conflict with our  static, spherically symmetric ansatz \eqref{metanz}, unless we consider a constant magnetic charge $g(r)=g_0$ or set $\alpha_1 = 0$. Similarly, the $t\phi$ and $r\theta$ components of the field equations do not vanish, the $\theta\theta$ and $\phi\phi$ components have angular dependence inconsistent with the ansatz 
\eqref{metanz}, and the $\phi$-component of 
\eqref{genmax} yields an equation for $
\nu(r)$ in terms of $g(r)$ 
unless we enforce $g(r)=g_0$ or set $\alpha_1 = 0$.  

In order to retain spherical symmetry and a non-trivial magnetic charge contribution we shall henceforth set  $\alpha_1 =0$. 
Using this, and defining a new combined dyonic charge ratio $x(r)=\frac{q(r)}{g(r)}g_0$, our radial metric function reduces to
\begin{equation}
\begin{aligned}
e^{-2\Lambda(r)}&=1-\frac{2m(r)}{r} - \frac{x(r)^2 r^2}{12 g_0^2 \alpha_2}\\
    &=1-\frac{2 \tilde{m}(r)}{r},
\end{aligned}
\end{equation}
where $\tilde{m}(r)$ is an effective mass that  includes a contribution from the dyonic charge term (namely, $\tilde{m}(r)=m(r)+\frac{x(r)^2 r^3}{24 g_0^2 \alpha_2}$) . The $r^2$ dependence of the dyonic term   means that in practice it acts like an effective cosmological constant on the spacetime. 

With these substitutions the pure QTEM version of the TOV equations (in terms of the mixed dyonic charge ratio function $x(r)$  and the physical mass $m(r)$) are:
\begin{equation}\label{eq:TOV}
\begin{aligned}
\frac{1}{\sqrt{\alpha_2}g_0} \frac{dx(r)}{dr} &= 4 \pi r^2\rho_x(r) e^{\Lambda (r)}   \\
\frac{d m(r)}{dr}&=4\pi r^2 \rho(r)-\frac{r^3 x(r) x'(r)}{12 g_0^2\alpha_2}\\
\frac{dp(r)}{dr}&=\frac{x(r) x'(r)}{16 \pi g_0^2 \alpha_2}+ e^{2 \Lambda(r)}\left(p(r)+\rho(r)\right) \left(\frac{r^3 x(r)^2-12 g_0^2 \alpha_2 \left(4\pi r^3 p(r)+m(r)\right)}{12 r^2 g_0^2 \alpha_2} \right)\\
\frac{d\nu(r)}{dr}&=\frac{1}{(p(r)+\rho(r))} \left( \frac{x(r) x'(r) }{16 \pi  g_0^2\alpha_2}-p'(r)\right)
\end{aligned}
\end{equation}
where $\sqrt{\alpha_2} \rho_x = \tilde{\rho}_x $ and the prime denotes a derivative with respect to $r$.

\subsection{Equation of State/Charge Distribution Model}\label{sec:eoscharge}

We employ a standard non-interacting quark matter equation of state 
\begin{equation}
    \rho(r) = 3 p(r) + 4 B_\mathrm{eff}
\end{equation}
which depends on the effective MIT bag constant $B_\mathrm{eff}$. We also generate solutions for two simple charge density models found in the literature \cite{zhang_2021_stellar,ray2003,arbanil2015,goncalves2020}. Our first charge distribution model (model A) assumes that dyonic charge density is directly proportional to mass density:
\begin{equation}\label{modA}
 \rho_x(r)=\zeta \rho(r).
\end{equation}
The second model for dyonic charge ratio function explicitly assumes that it scales with spatial volume, namely 
\begin{equation}\label{modB}
    x(r)=X\left(\frac{r}{R}\right)^3=\beta r^3,
\end{equation}
where $X=x(R_{L/S})$ represents the total dyonic charge ratio at the large (including negative pressure shell) and small (only the positive inner core) stars' surfaces respectively.

\subsection{Units and Scaling}\label{sec:units}

Before solving the equations we first recast them into a unitless form \cite{zhang_2021_unified,zhang_2021_stellar}, using the effective MIT bag constant $B_\mathrm{eff}$ as the fundamental scale. In the results we assume a typical bag constant value $B_\mathrm{eff}=60 \;\mathrm{MeV/fm^3}$ \cite{arbanil2015} on the unitful axes, however the following redefinitions can be used to convert the unitless results for any desired value of $B_\mathrm{eff}$ \cite{zhang_2021_unified}. First, relevant EOS parameters can be rescaled like:

\begin{equation}\label{eq:rescale_prho}
    \bar{p}=\frac{p}{4B_\mathrm{eff}},\; \bar{\rho}=\frac{\rho}{4B_\mathrm{eff}}
\end{equation}

leaving us with a unitless non-interacting quark equation of state:

\begin{equation}\label{eq:EOS_unitless}
    \bar{\rho}(\bar{r}) = 3 \bar{p}(\bar{r}) + 1.
\end{equation}

Furthermore we can write
\begin{equation}\label{eq:rescale_mr}  \bar{m}=m\sqrt{4B_\mathrm{eff}} \qquad \bar{r}=r\sqrt{4B_\mathrm{eff}}
\end{equation} 
along with the charges
\begin{equation}\label{eq:rescale_qrho} 
\bar{q}=q\sqrt{4 B_\mathrm{eff}} \qquad \bar{g}=g\sqrt{4B_\mathrm{eff}}\qquad \bar{x}=x\sqrt{4B_\mathrm{eff}},
\end{equation}
the charge density constants
\begin{equation}\label{eq:rescale_zetabeta} \bar{\zeta}=\frac{\zeta}{4 B_\mathrm{eff}} \qquad  \bar{\beta}=\frac{\beta }{ 4 B_\mathrm{eff}},
\end{equation}
and
\begin{equation}\label{eq:rescale_galpha} \bar{\rho}_x=\frac{\rho_x}{(4 B_\mathrm{eff})^2} \qquad \bar{\alpha}_2=\alpha_2 \cdot 4B_\mathrm{eff}.
\end{equation}
  With these substitutions the TOV equations modified for QTEM \eqref{eq:TOV} can be cast in unitless form by simply replacing the unbarred symbols with their barred counterparts.

\section{Results}\label{sec:results}

Noting that all equations in 
\eqref{eq:TOV} depend only on
$\bar{x}(\bar{r})/(\bar{g}_0\sqrt{\bar{\alpha_2}})$, we set $\bar{\alpha}_2 \bar{g}_0^{2} = 1$
without loss of generality, and plot most of the results over the total physical mass $M$. For both charge models we find a second solution branch starting at large mass and radius. This is a result of the exotic QTEM pressure profiles that turn upward again after crossing zero for the first time (see figure \ref{fig:modelApressures}). 
 These are solutions that contain a `normal' quark star core (from the lower mass branch) inside an outer envelope (or shell) of negative pressure (but positive density). The inner `core' consists of matter with positive density and pressure with the pressure vanishing at the inner shell boundary.    This phenomenon is a manifestation of the $r^2$ dependence of the interior metric function
\eqref{Lamanz} (inherited from the vacuum metric \eqref{vacanz}), corresponding to a type of vacuum tension analogous to that of a positive cosmological constant.

Given the energy conditions on a perfect fluid ($\bar{\rho}+\bar{p}\geq0$, $\bar{\rho}\geq0$, $\bar{\rho}\geq |\bar{p}|$, $\bar{\rho}+3\bar{p}\geq0$) \cite{Maeda_2022} along with our equation of state \eqref{eq:EOS_unitless} it is straightforward to show that these are all satisfied as long as $\bar{p}\geq-4$,  which is the case 
for all solutions we have obtained.

\subsection{Charge Model A}\label{sec:resultsa}

\begin{figure}
        \subfloat[\label{fig:modelAMRsmallzetazoomedin}]{
        \includegraphics[width=7cm]{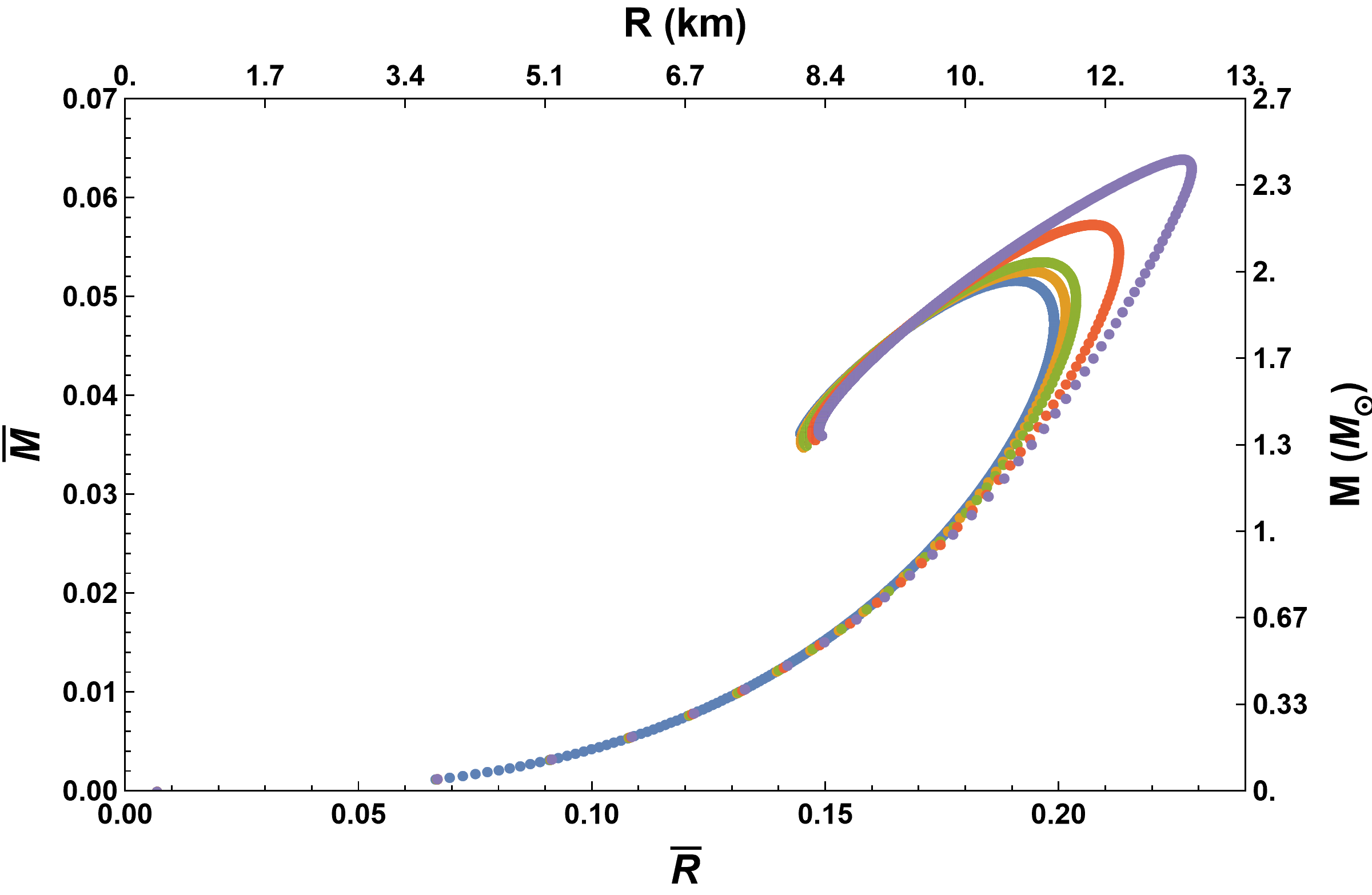}
        }\hfill
        \subfloat[\label{fig:modelAMRhosmallzetazoomedin}]{
        \includegraphics[width=8cm]{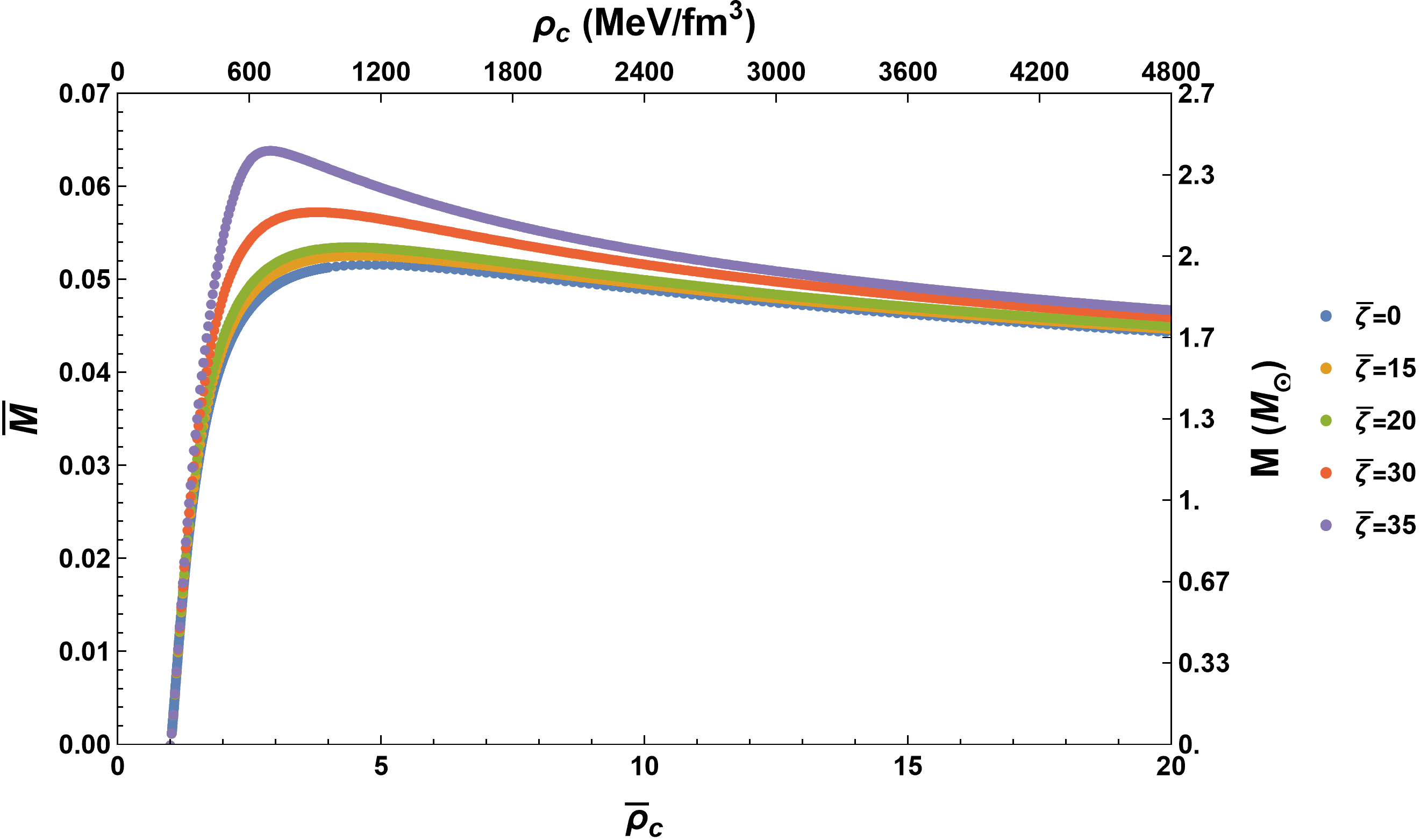}
        }        

	\caption[]{The `normal' set of solutions for charge model A with small $\bar{\zeta}$. We see the classic characteristic $M/R$ `hook' shape that is usually present for quark matter \cite{glendenning_1997_compact,zhang_2021_unified,zhang_2021_stellar,banerjee_2021_strange,banerjee_2021_quark,gammon_2024,gammon_2025_chargedquarkstars4degb}.  Adding a dyonic charge density that is small (meaning electric charge density is small when scaled by magnetic charge density) inflates the mass-radius profiles seen here.}
	\label{fig:modelAsmallzetazoomedin}
\end{figure}

For various smaller values of $\bar{\zeta}$ (representing small electric charge density scaled by magnetic charge density) in model A \eqref{modA} we obtain mass-radius curves that are qualitatively the same as the hook shapes typically observed for $M/R$ curves resulting from the equation of state for (un)charged quark matter \cite{Weber_2012,glendenning_1997_compact,zhang_2021_stellar,gammon_2024,gammon_2025_chargedquarkstars4degb}. This is shown in the left panel of figure~\ref{fig:modelAsmallzetazoomedin}. Plots of mass as a function of central density $\rho_c$ shown in the right panel that correspond to each of the curves at left are likewise qualitatively similar.

\begin{figure}
        \subfloat[\label{fig:modelAMRsmallzetazoomedout}]{
        \includegraphics[width=7cm]{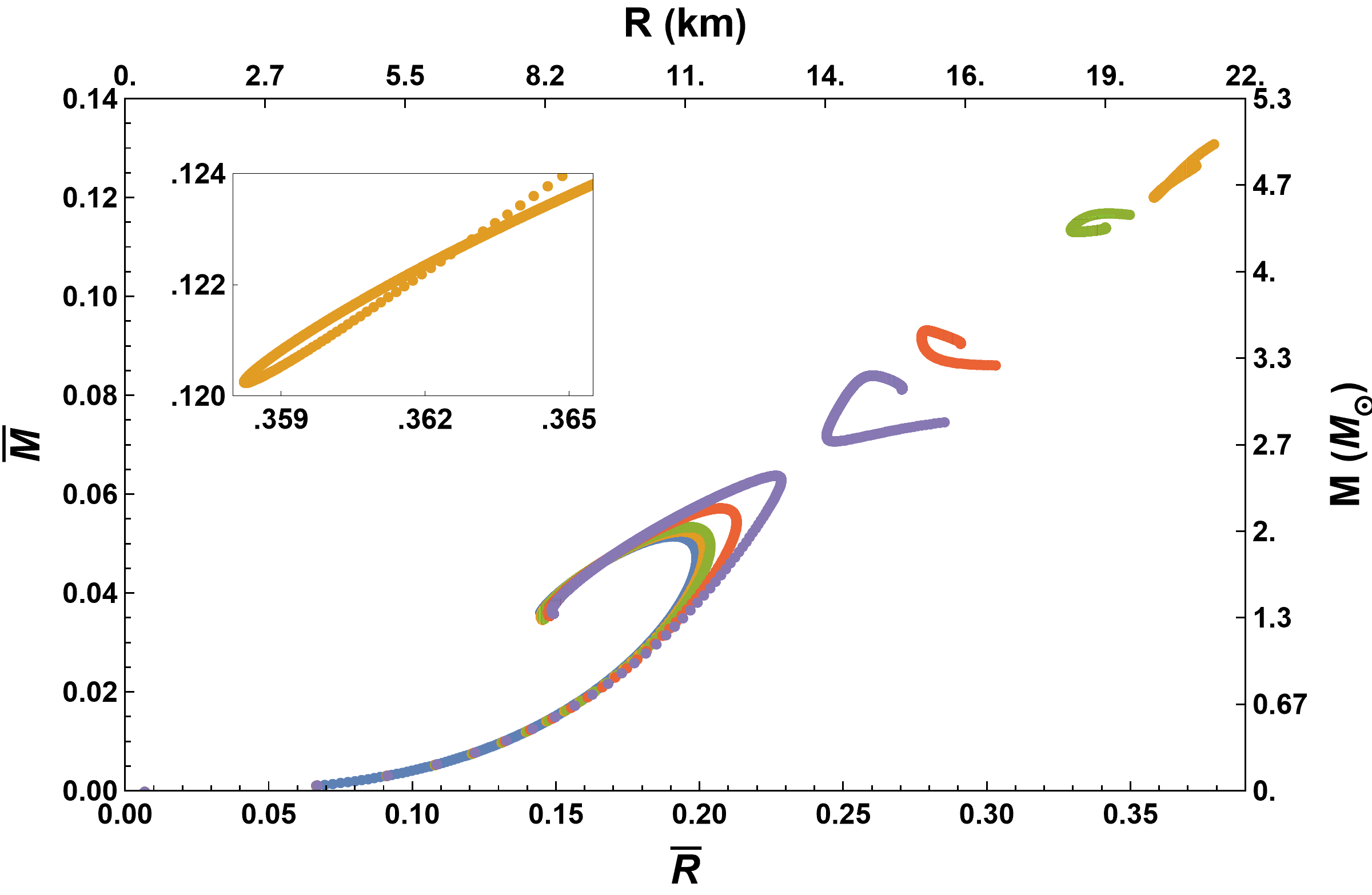}
        }\hfill
        \subfloat[\label{fig:modelAMRhosmallzetazoomedout}]{
        \includegraphics[width=8cm]{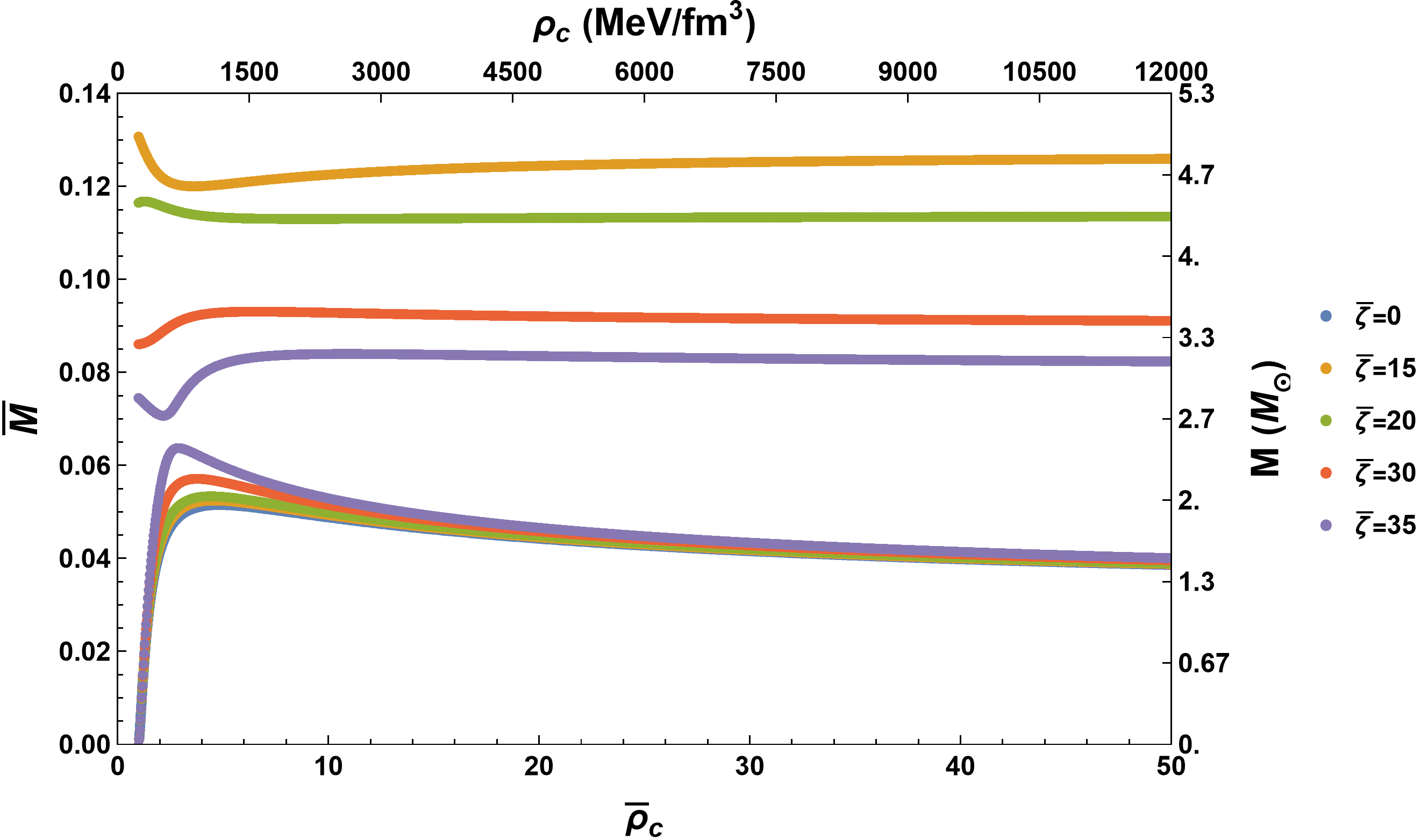}
        }        

	\caption[]{Zooming out from figure~\ref{fig:modelAsmallzetazoomedin} we see that a whole other solution branch exists if we consider another stellar surface to be at the location where the pressure function crosses zero from below (see figure \ref{fig:modelApressures}). As $\bar{\zeta}$ increases, these branches start to move toward one another.}
\label{fig:modelAzoomout}
\end{figure}

\begin{figure}
        \subfloat[$\bar{\zeta}=0$. \label{fig:modelApressurezeta0}]{ 
        \includegraphics[width=7.6cm]{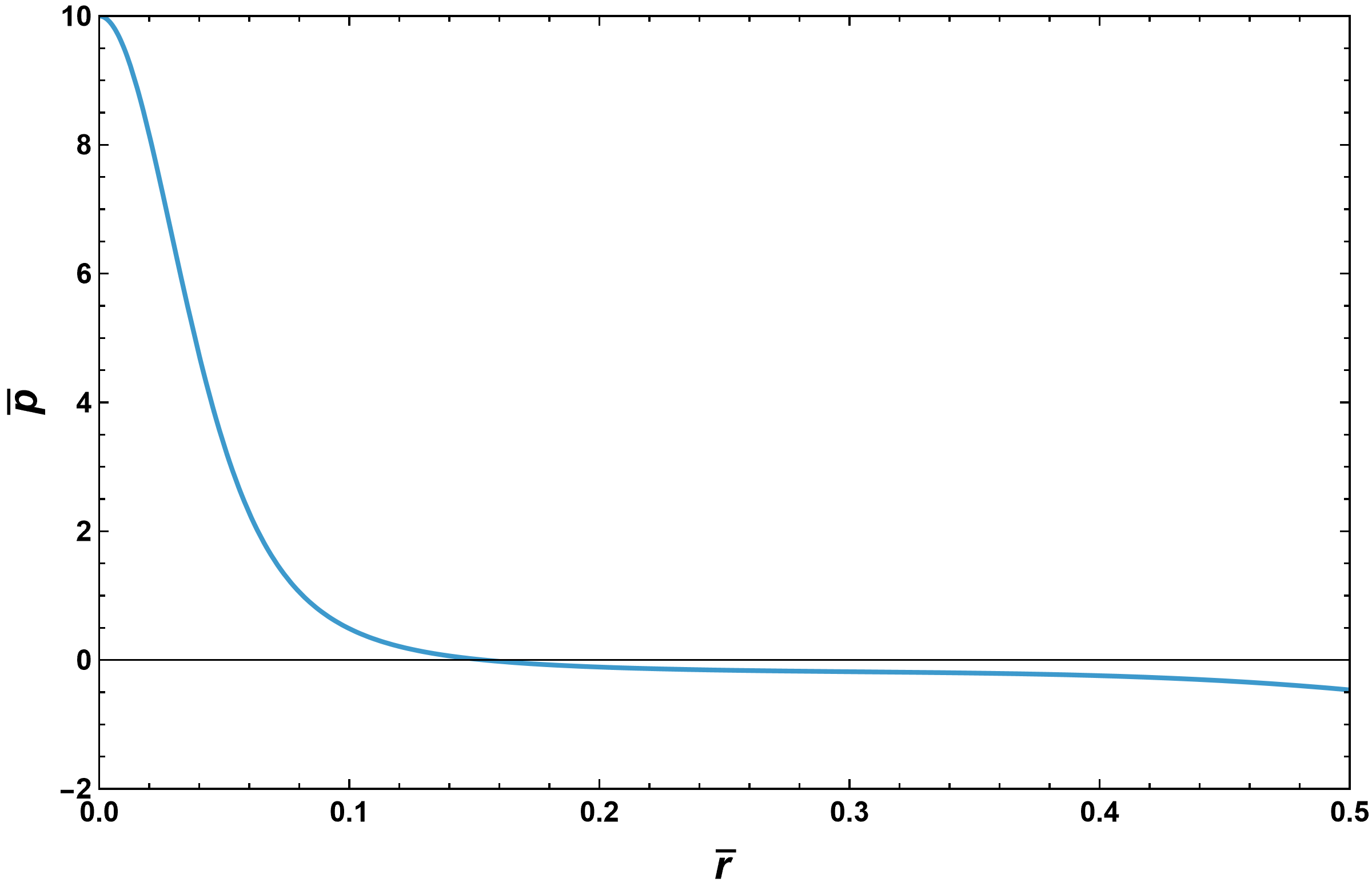} 
        }\hfill
        \subfloat[$\bar{\zeta}=30$ \label{fig:modelApressurezeta30}]{
        \includegraphics[width=7.6cm]{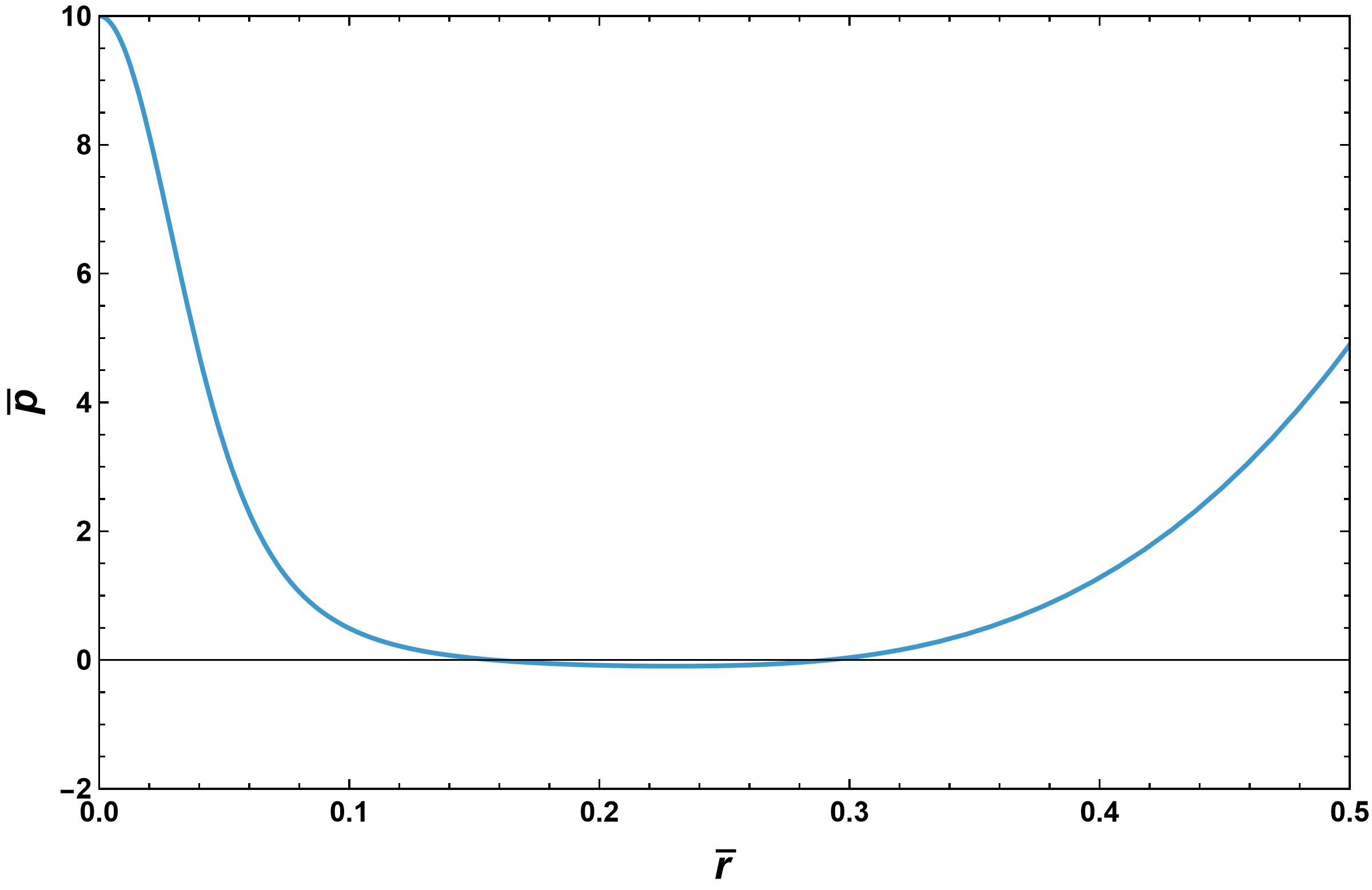}
        }        

	\caption[]{Comparison of the pressure profiles for the $\bar{\zeta}=0$ and $\bar{\zeta}=30$ cases of charge model A. Non-zero dyonic charge density causes the pressure profile to turn upward outside of the `primary' star's surface before crossing 0 again from below, forming a surrounding shell of matter with negative pressure which is held together by tension. Because of this the pressure profiles for dyonic quasi-topological quark stars have two roots, with the second manifesting as the upper branch in the $M/R$ diagrams.}
	\label{fig:modelApressures}
\end{figure}

 However  plots of $p(r)$ exhibit qualitatively different behaviour than those of a normal electrically charged quark star. A typical example is shown in figure~\ref{fig:modelApressures}, where  $\bar{\zeta}=30$ is compared to the uncharged case.  In the latter case (left panel~\ref{fig:modelApressurezeta0}) the function $p(r)$ has a single root $r=R$, whose value
 defines the radius of the star.  However in the dyonically charged case (right panel~\ref{fig:modelApressurezeta30}), $p(r)$ attains a local minimum, after which it grows without bound.  If
 this minimum occurs at $p<0$ then there are two possible radii for the dyonic star.   The smaller root $r=R_S$ corresponds to a star of smaller mass  and radius,  whereas the larger root $r=R_L$ corresponds to a star of larger mass and radius.  This latter case corresponds to a positive mass star with an outer shell held together by tension (negative pressure) and an inner core of positive pressure. In this way the branch 2 stars each contain a branch 1 star inside of their negative pressure shells.  The boundary between the core and the shell is at the first root of the pressure function.  
 These additional (branch 2) solutions are shown in figure~\ref{fig:modelAzoomout}, with the left panel~\ref{fig:modelAMRsmallzetazoomedout} showing the mass-radius relation and 
 the right panel~\ref{fig:modelAMRhosmallzetazoomedout} depicting the mass as a function of central density. The smaller values
 of $\bar{\zeta}$ produce $M/R$ plots which exhibit a `fish' structure in the upper branch, shown in the inset of the left panel~\ref{fig:modelAMRsmallzetazoomedout}.

\begin{figure}
    \centering
    \includegraphics[width=0.5\linewidth]{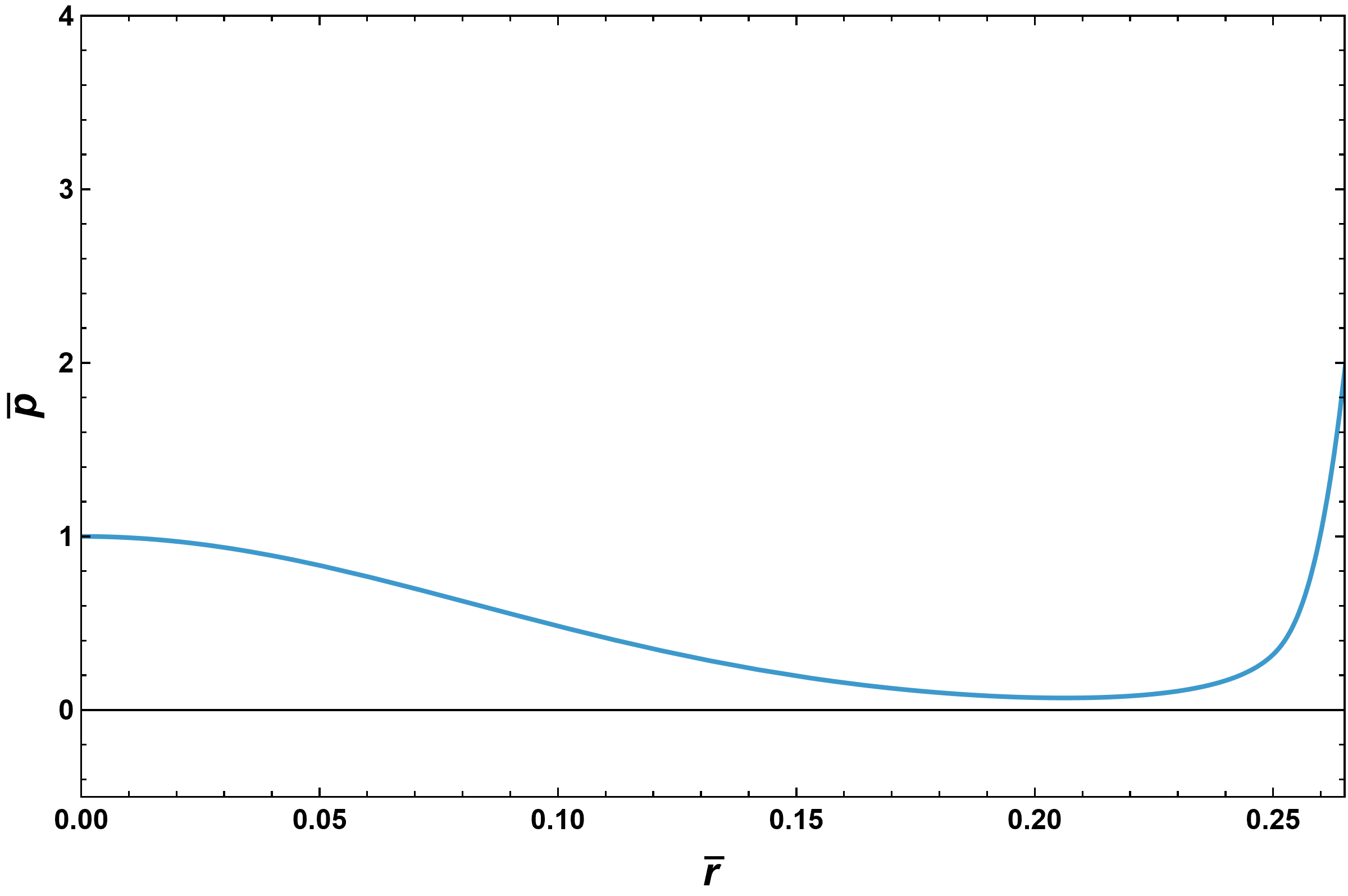}
    \caption{A plot of an sample pressure profile for $\bar{\zeta}=50$ and $\bar{p}_0=1$, which lies in the part of parameter space with no quark star solutions. We see the reason for this behaviour - the pressure minima is above zero and thus the function has no roots.}
    \label{fig:modelApressureprofilezeta50}
\end{figure}

 
These disparate branches eventually merge for large enough $\bar{\zeta}$, which results in a range of central densities for which there are no solutions.
 In such cases the local minimum of the  pressure profile is always positive, never crossing 0.  An example for $\bar{\zeta}=50$ is shown in figure~\ref{fig:modelApressureprofilezeta50}.
 
\begin{figure}
    \centering
    \includegraphics[width=0.5\linewidth]{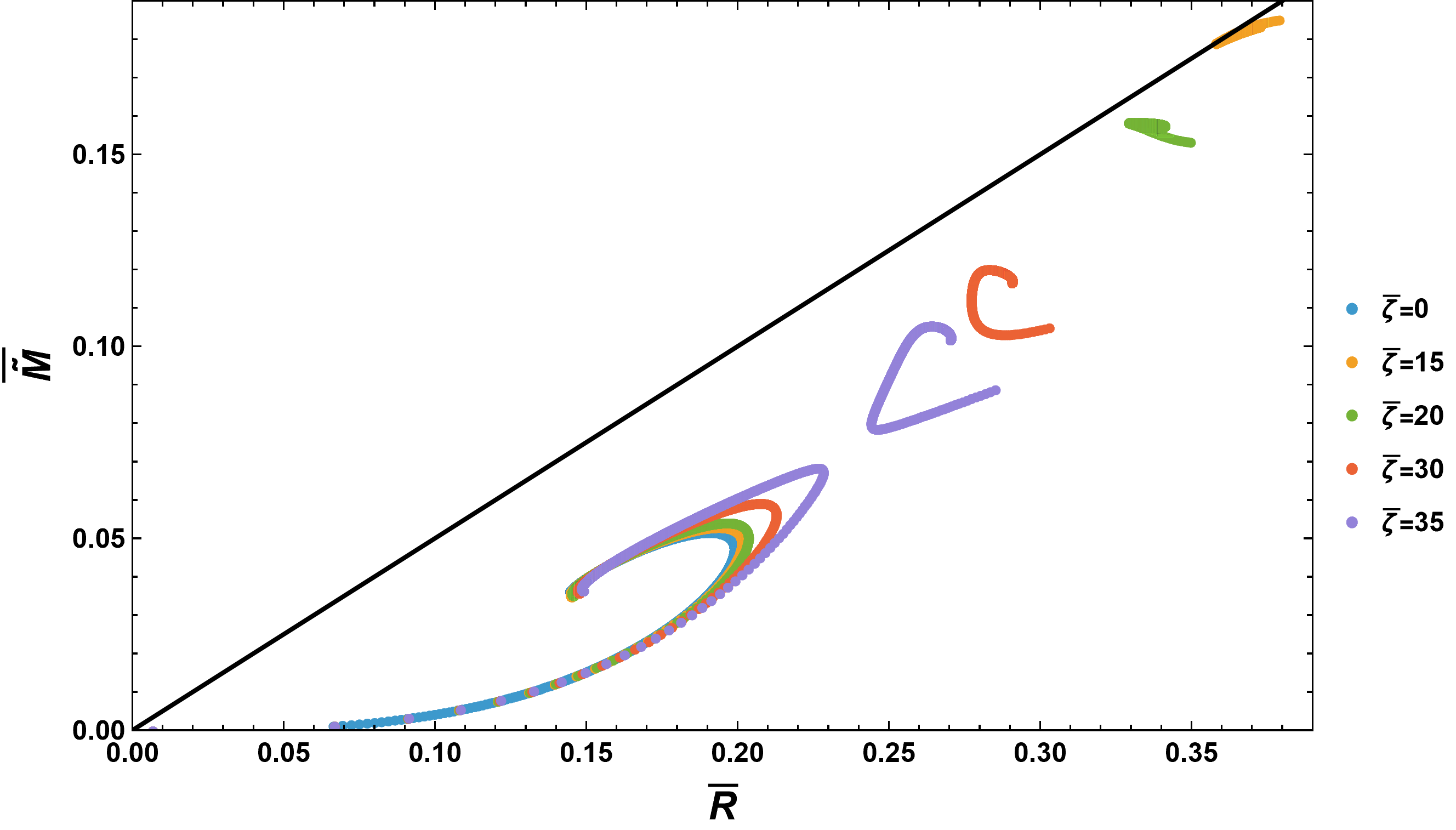}
    \caption{Unitless effective mass $\bar{\tilde{M}}$ over radius $\bar{R}$ for the same five values of $\bar{\zeta}$ plotted in figure \ref{fig:modelAMRsmallzetazoomedout}, alongside the corresponding dyonic black hole horizon ($\tilde{M} = \frac{R}{2}$). We see why for $\bar{\zeta} < 15$ the solution for $p(r)$ becomes complex.}\label{fig:modelAMRsmallzetaeffectivemass}
\end{figure}

 Another interesting feature of the charge model A results is that the second branch solutions  have pressure profiles that can become complex due black hole horizon formation before the second root is ever reached (provided $\bar{\zeta}$ is small enough). We have found that the smallest $\bar{\zeta}$ for which none of the solutions have horizon formation is roughly $\bar{\zeta}=15$, whose solutions are shown in figure~\ref{fig:modelAMRsmallzetaeffectivemass} alongside the corresponding dyonic black hole horizon.  

\begin{figure}
        \subfloat[\label{fig:modelAMRtransitionzeta}]{
        \includegraphics[width=7cm]{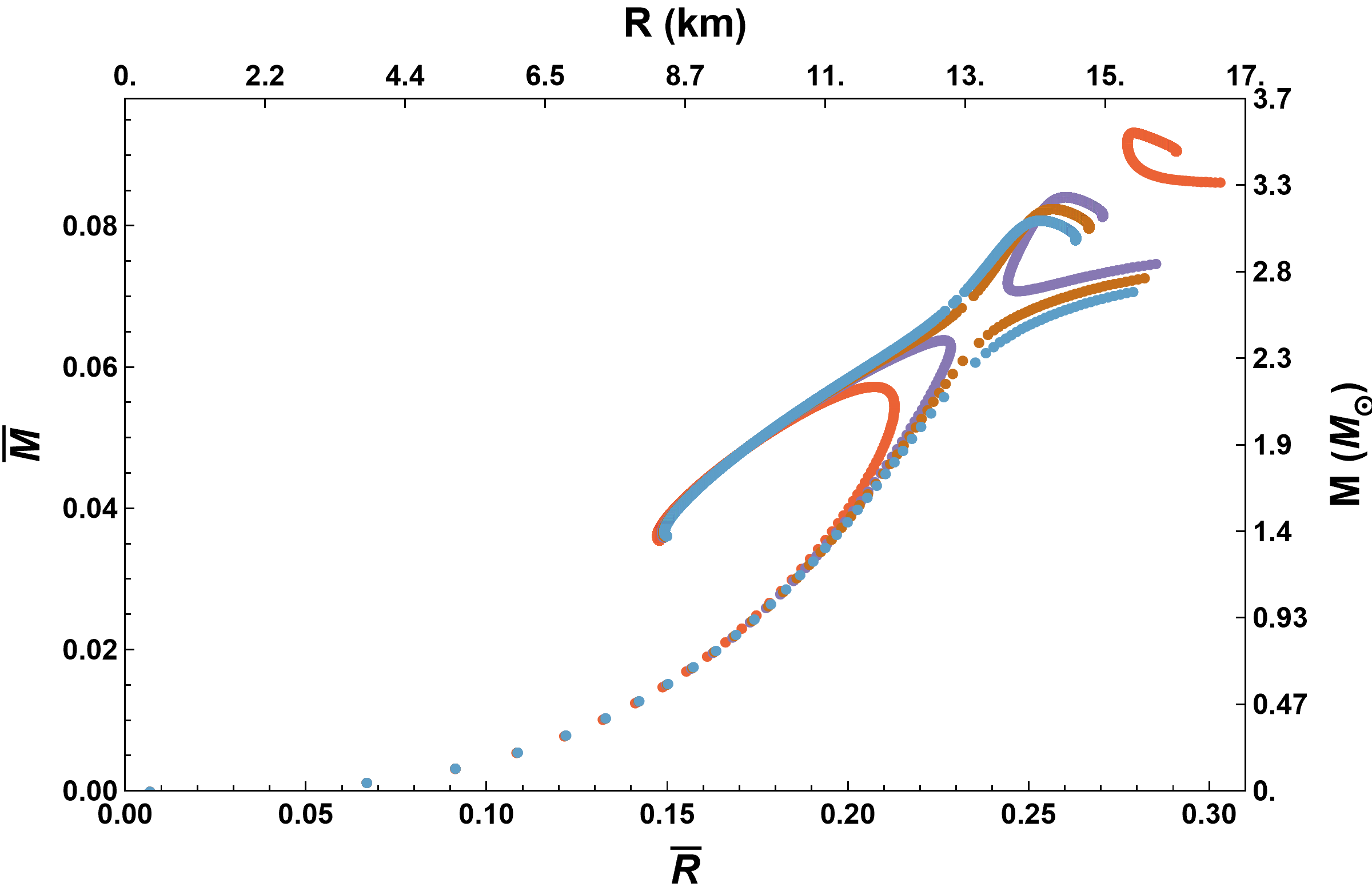}
        }\hfill
        \subfloat[\label{fig:modelAMRhotransitionzeta}]{
        \includegraphics[width=8cm]{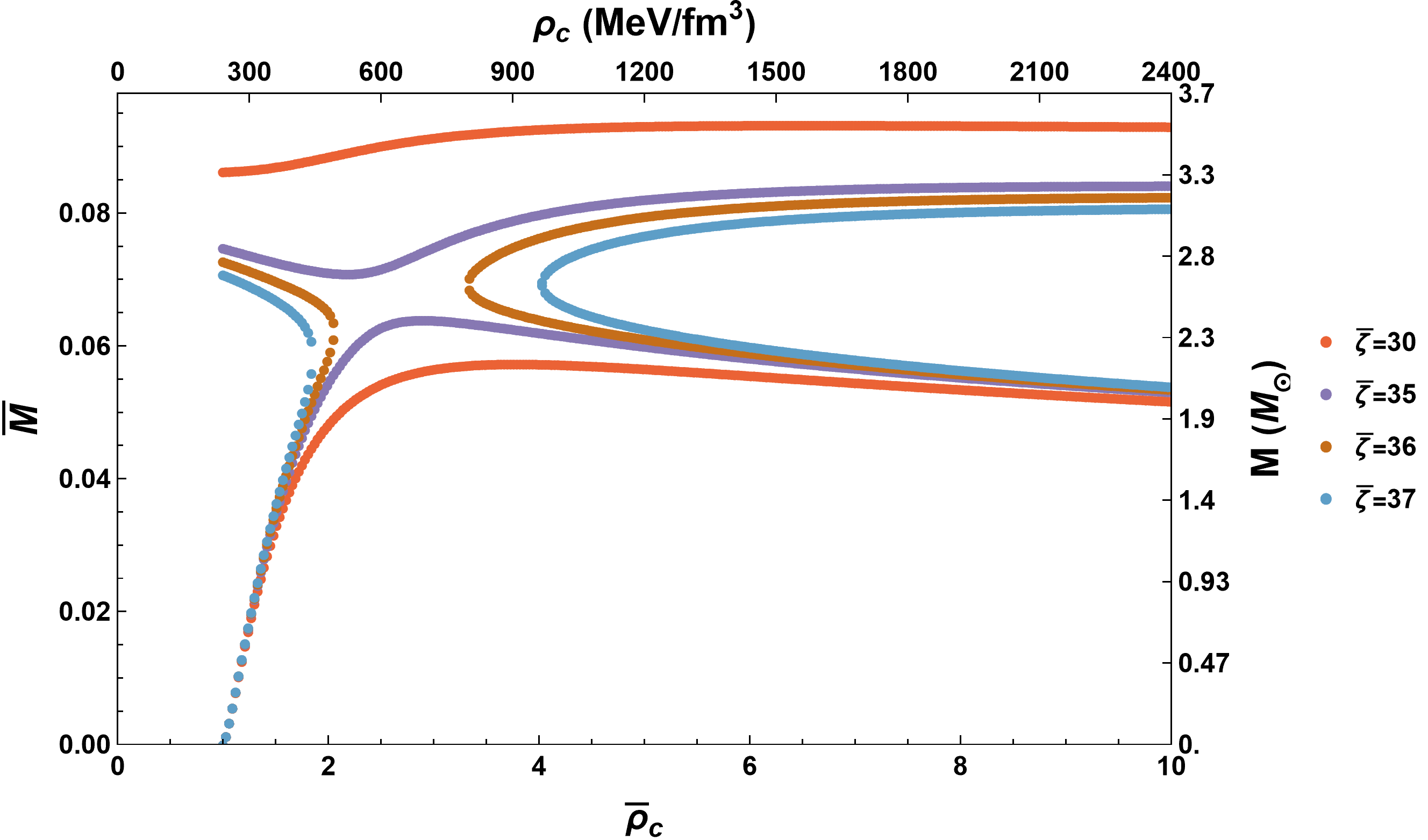}
        }        
	\caption[]{These are the $\bar{\zeta}$ values over which the two separated solution branches in the $M/R$ plot join together, effectively defining a range of central pressures (and central densities) for which there are no solutions in the $M/\rho$ plot (with the first example above being $\bar{\zeta}=36$). This is due to the local minima of $p(r)$ (which normally separates the two roots) being above the x-axis (ie. the pressure is always positive), as shown in figure~\ref{fig:modelApressureprofilezeta50}.}
\label{fig:modelAmerge}	
\end{figure}
\begin{figure}    \subfloat[\label{fig:modelAMRlargezeta}]{
        \includegraphics[width=7cm]{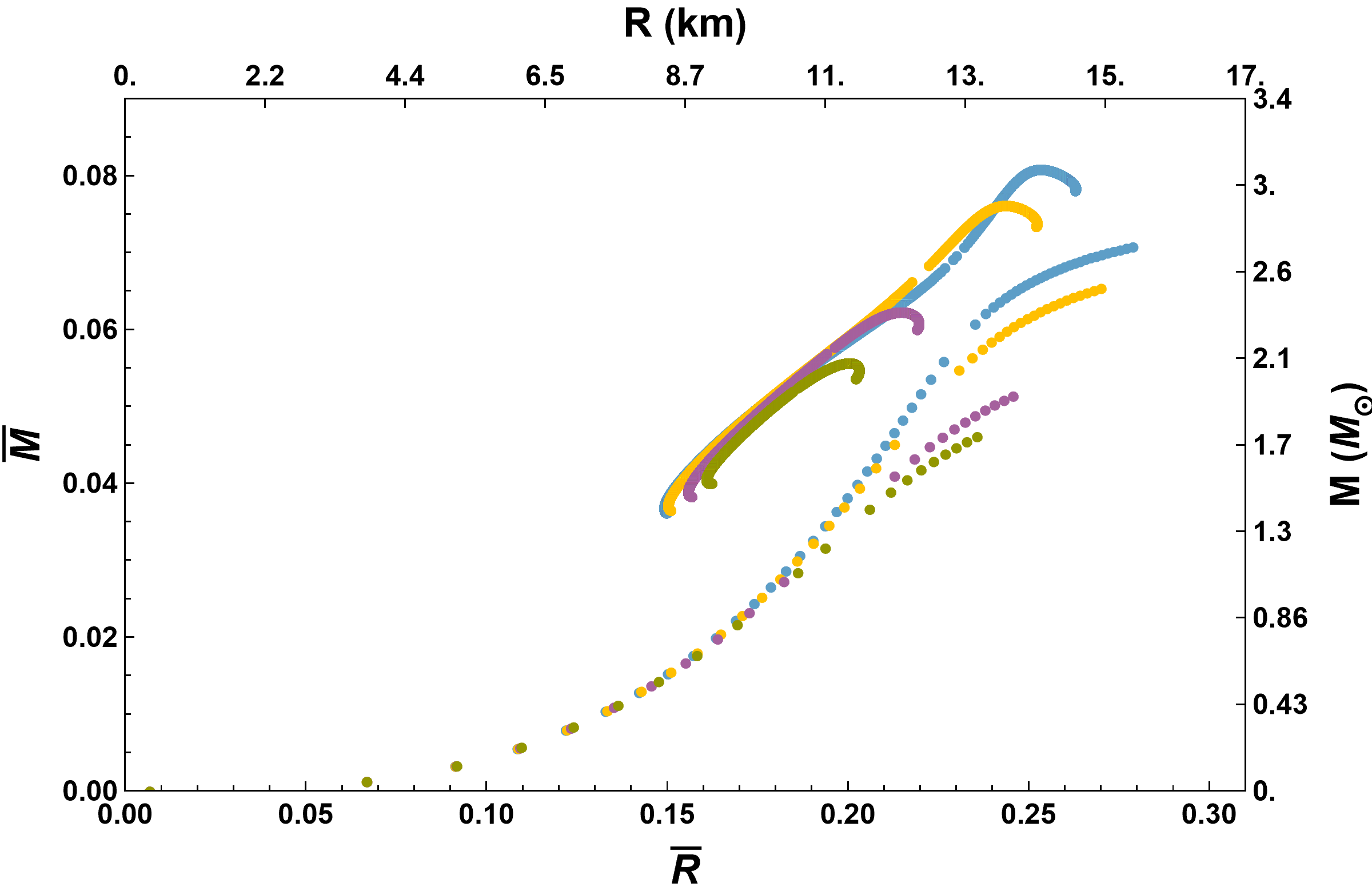}
        }\hfill
        \subfloat[\label{fig:modelAMRholargezeta}]{
        \includegraphics[width=8cm]{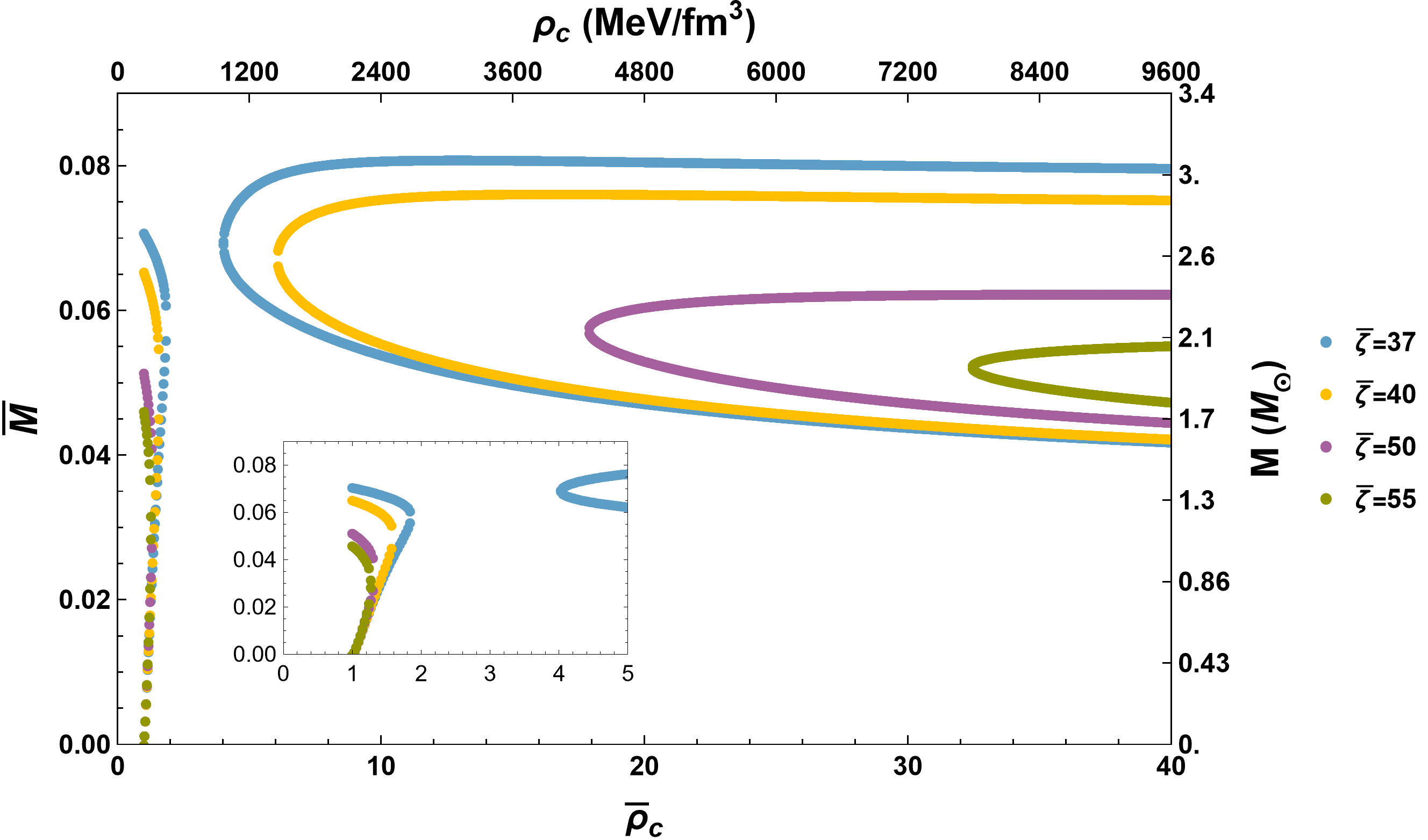}
        }        

	\caption[]{For larger $\bar{\zeta}$ the higher-central density parts of the solution curve form the characteristic `paperclip' shapes higher in the $M/R$ plots, whereas the low-density part makes up the longer curves starting from the origin, looking somewhat like the bottom part of a normal quark matter `hook' from the $\bar{\zeta}=0$ case.}
\end{figure}
For larger $\bar{\zeta}$ another interesting phenomenon occurs: the two solution branches merge. This takes place for $35< \bar{\zeta} < 36$, as shown
in figure~\ref{fig:modelAmerge}, with the left panel~\ref{fig:modelAMRtransitionzeta} showing the mass-radius curves and the right panel~\ref{fig:modelAMRhotransitionzeta} showing the mass as a function of central density. In $M/\rho_c$ space we see this manifesting as a range of central densities for which there are no solutions. It should be noted that all of the upper branch solutions plotted are less compact than the corresponding black hole solution.

\FloatBarrier

\subsection{Charge Model B}\label{sec:resultsb}

\begin{figure}
        \subfloat[\label{fig:modelBMRsmallzetazoomedin}]{
        \includegraphics[width=7cm]{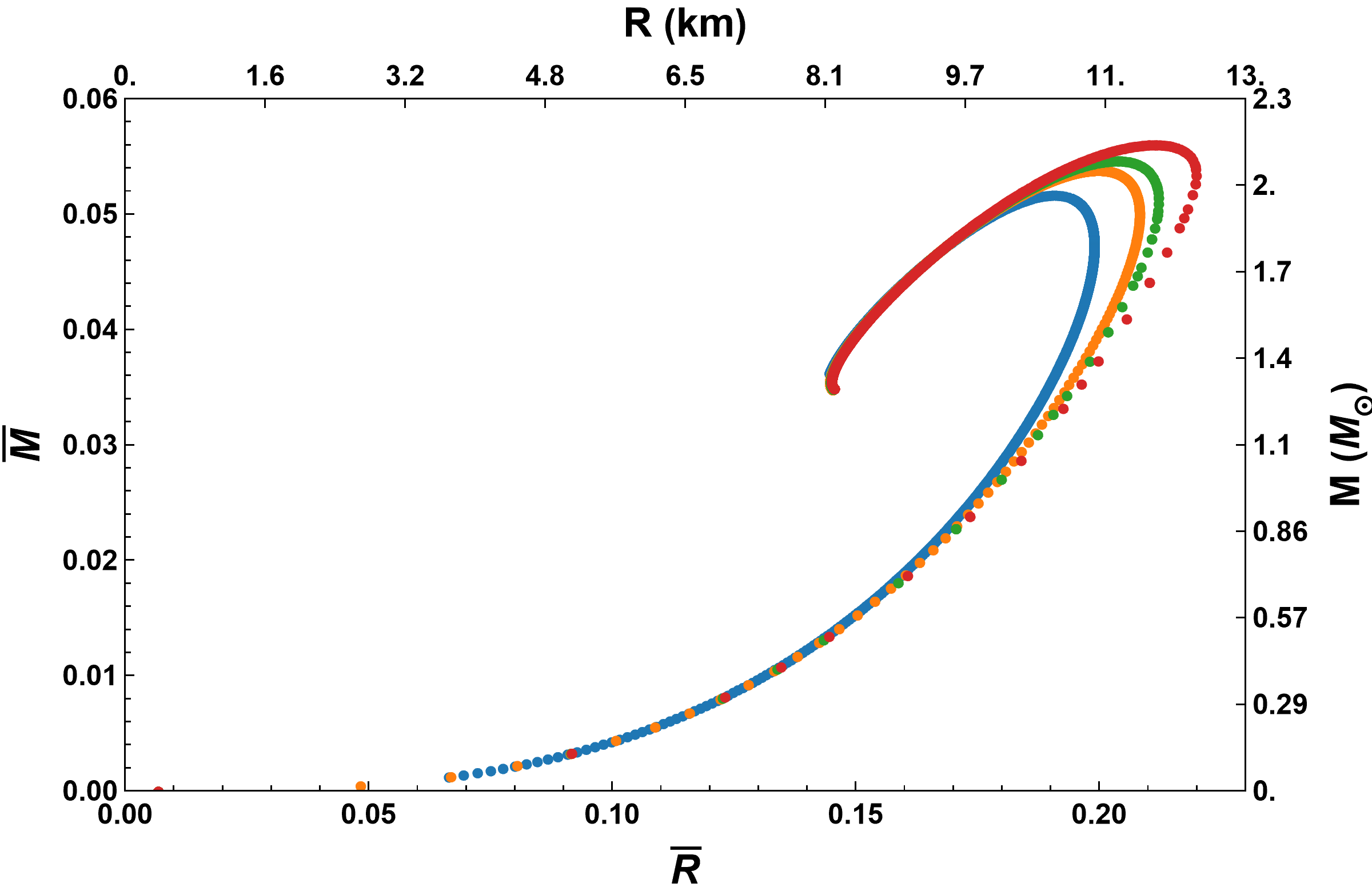}
        }\hfill
        \subfloat[\label{fig:modelBMRhosmallbetazoomedin}]{
        \includegraphics[width=8cm]{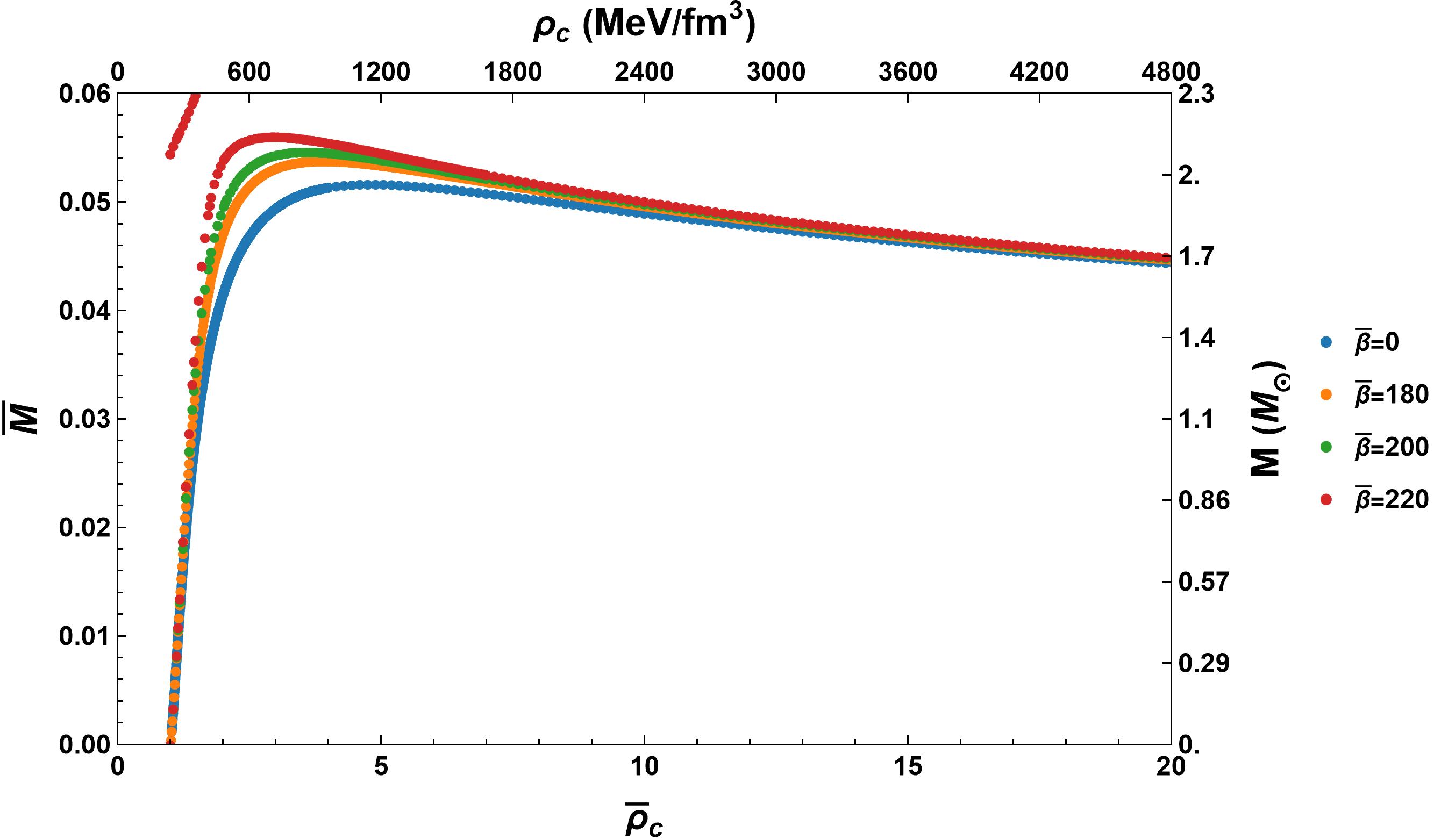}
        }        

	\caption[]{The `normal' set of solutions for charge model B with small $\bar{\beta}$.}
	\label{fig:modelBsmallbetazoomedin}
\end{figure}

\begin{figure}
        \subfloat[\label{fig:modelBMRsmallbetazoomedout}]{
        \includegraphics[width=7cm]{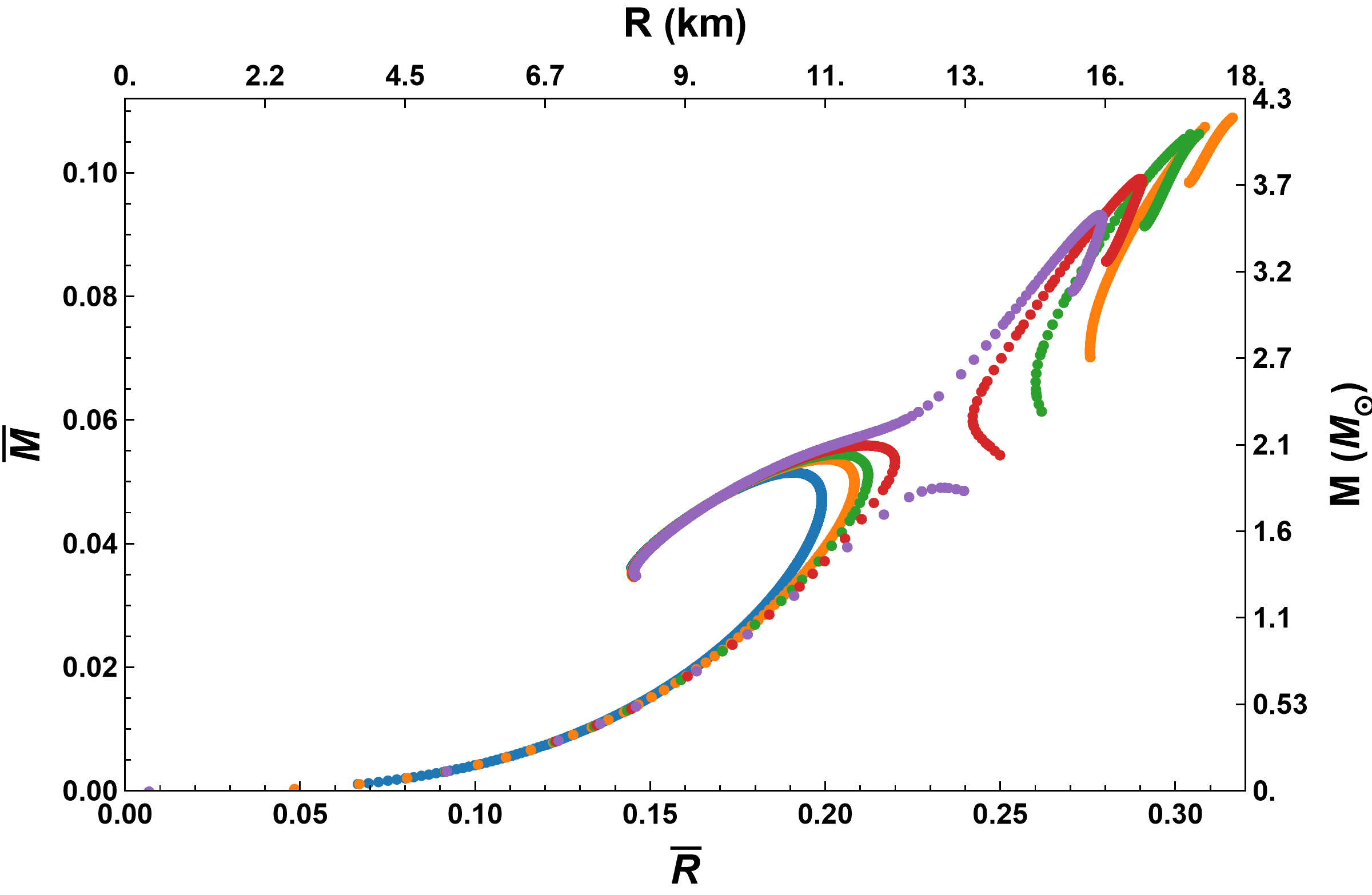}
        }\hfill
        \subfloat[\label{fig:modelBMRhosmallbetazoomedout}]{
        \includegraphics[width=8cm]{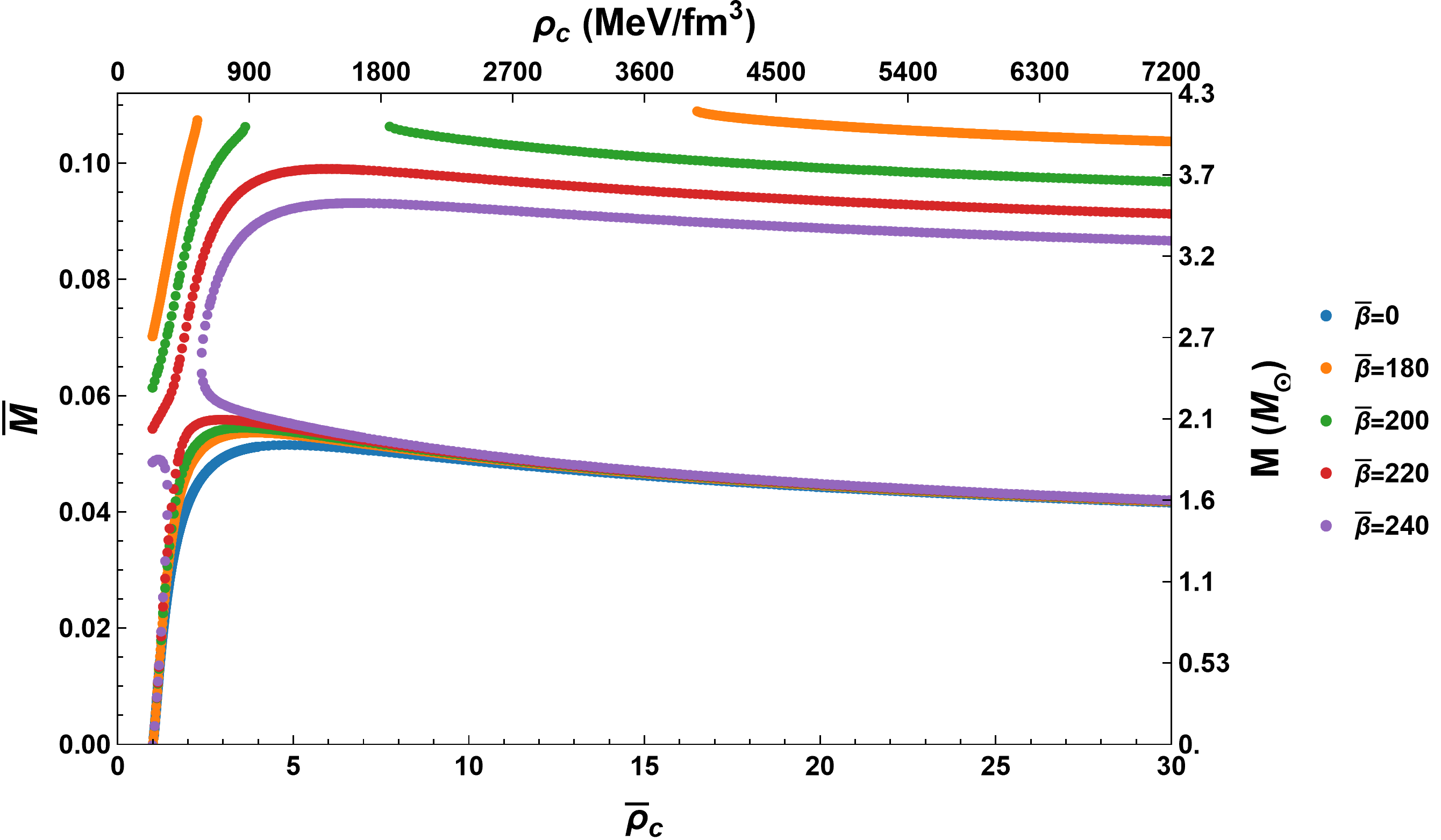}
        }        

	\caption[]{Zooming out from \ref{fig:modelBsmallbetazoomedin} we see that just like in charge model A a whole other solution branch exists if we consider another stellar surface to be at the location where the pressure function crosses zero from below.  Unlike model A, the \textit{small} $\bar{\beta}$ solutions in $M/\rho_c$ space also have gaps for a small range of central pressures. In these cases the pressure curve actually turns back around before reaching a second root from below. When $\bar{\beta}$ gets large enough these two branches eventually merge like the model A case (as can be seen with the $\bar{\beta}=240$ case above). This again results in a range of central pressures with no quark star solutions. These merger gaps are notably different from the gaps in the smallest $\bar{\beta}$ curves, resulting from the entire pressure profile being positive ($p(r)>0$) rather than the turnaround behaviour discussed previously.}\label{fig:modelBsmallbetazoomedout}
\end{figure}

Our results for charge model B share many qualitative features with the results from charge model A. Introducing a small, non-zero dyonic charge parameter (in this case $\beta$) yields, in addition to standard quark star `hook' solutions (shown in figure ~\ref{fig:modelBsmallbetazoomedin}), the same exotic pressure profiles from section \ref{sec:resultsa} where a shell of negative pressure (but positive density) is present at the outermost part of the star. These  solutions with an outer envelope enclosing the positive pressure quark matter core form the second solution branch, shown in the top right part of figure~\ref{fig:modelBMRsmallbetazoomedout}. These have a notably  different shape compared to their model A counterparts (figure \ref{fig:modelAMRsmallzetazoomedout}).  Another unique feature is that below $\bar{\beta}\approx 200$ the upper solutions have gaps for some range of central pressures. In these cases the pressure solution curves turn back around before ever reaching the second root, diverging to negative infinity.

\begin{figure}
        \subfloat[\label{fig:modelBMRtransitionbeta}]{
        \includegraphics[width=7cm]{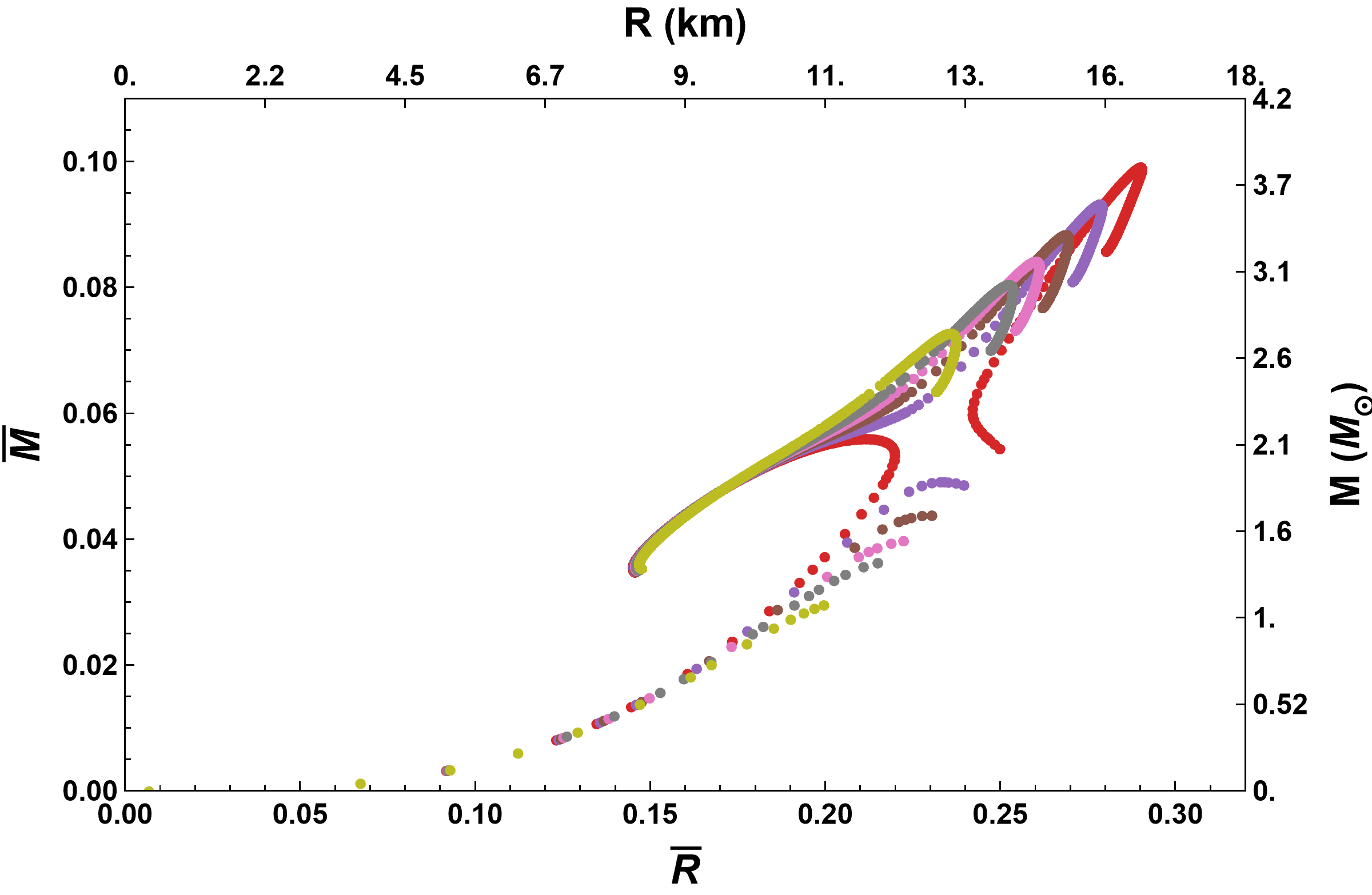}
        }\hfill
        \subfloat[\label{fig:modelBMRhotransitionbeta}]{
        \includegraphics[width=8cm]{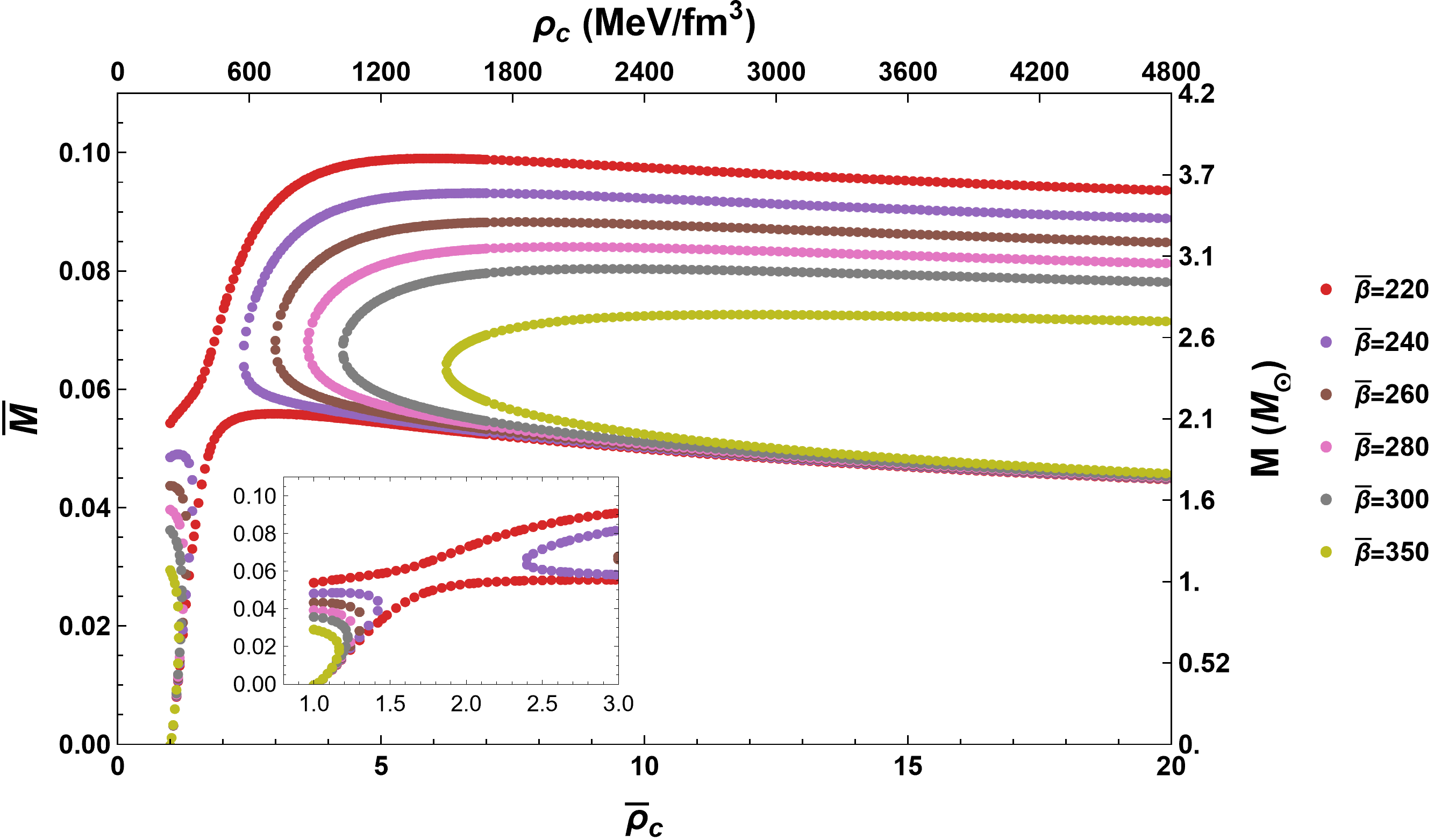}
        }        

	\caption[]{These are the $\bar{\beta}$ values over which the two separated solution groups in the $M/R$ plot join together, effectively defining a range of central pressures for which there are no solutions in the $M/\rho_c$ plot. This is again due to the local minima of $p(r)$ (which normally separates the two roots) being above the x-axis (meaning the pressure is always positive).}\label{fig:modelBtransitionbeta}
\end{figure}

\begin{figure}
        \subfloat[\label{fig:modelBMRlargebeta}]{
        \includegraphics[width=7cm]{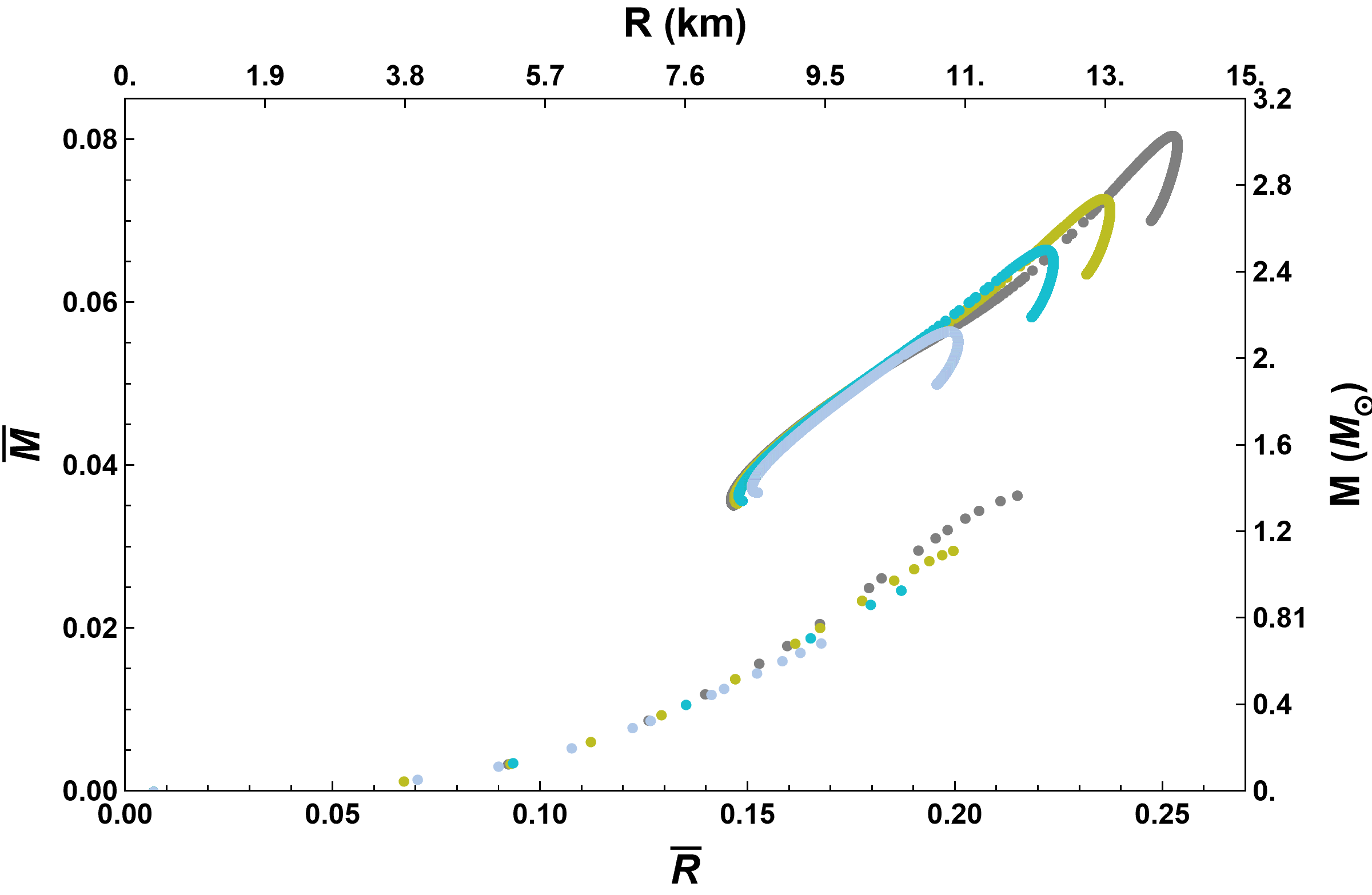}
        }\hfill
        \subfloat[\label{fig:modelBMRholargebeta}]{
        \includegraphics[width=8cm]{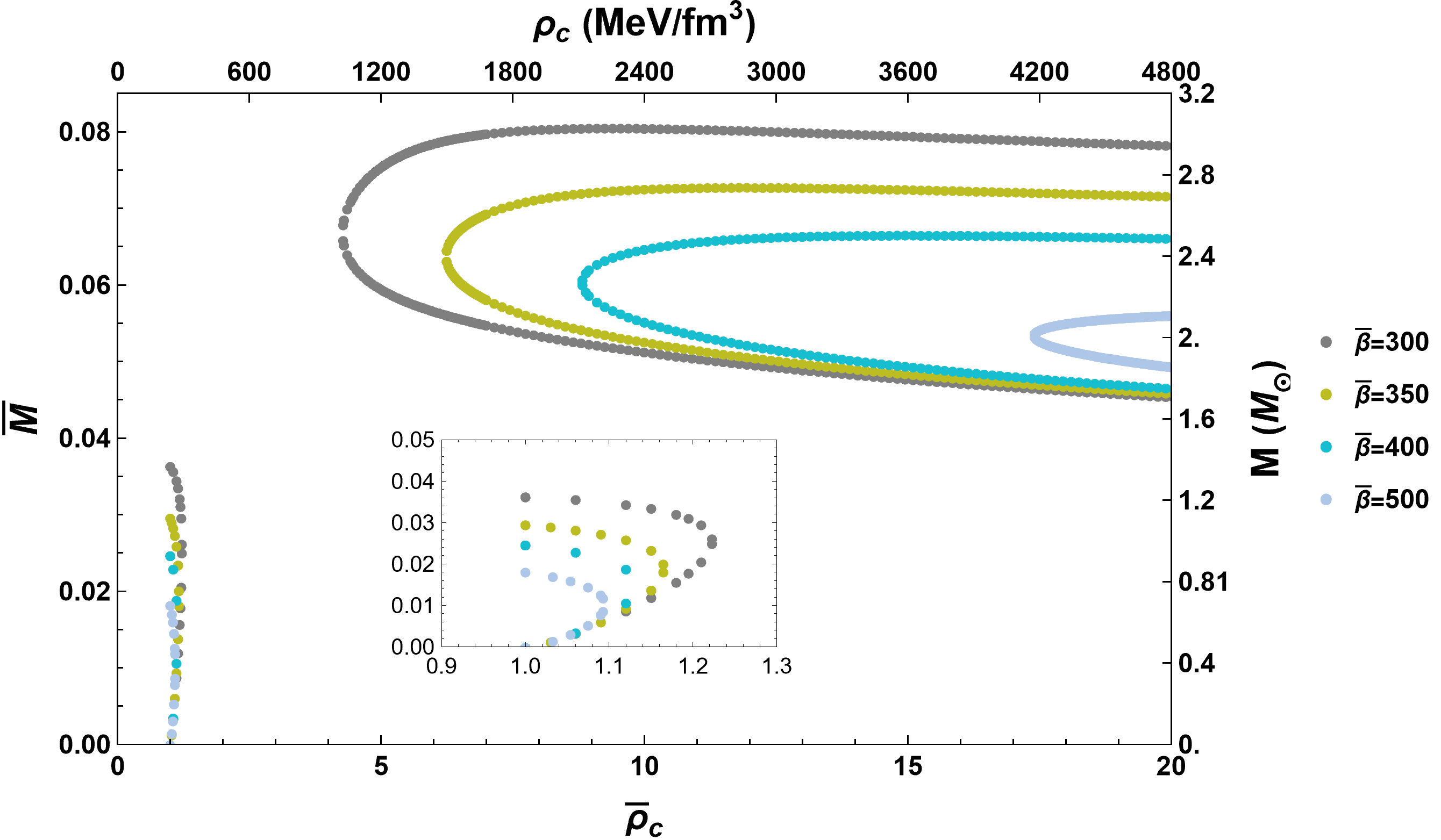}
        }        
	\caption[]{For larger $\bar{\beta}$ the higher-central density parts of the solution curve form the `paperclip' shapes higher in the $M/R$ plots, whereas the low-density part makes up the longer curves starting at the origin that look somewhat like the bottom part of a normal quark matter `hook' from the $\bar{\beta}=0$ case.}\label{fig:modelBlargebeta}
\end{figure}

As $\bar{\beta}$ further increases the branches merge together, once again manifesting as a range of central densities with no quark star solutions (figures \ref{fig:modelBMRhosmallbetazoomedout} and \ref{fig:modelBMRhotransitionbeta}). As $\bar{\beta}$ continues to increase (see figure \ref{fig:modelBlargebeta}) we see the upper part of the $M/R$ curve (representing the higher central density solutions) eventually settle into a similar `paperclip' shape that was also seen in the large $\bar{\zeta}$ limit of model A. Once again, all presented solutions are less compact than the corresponding dyonic QTEM black hole.  With this, it seems that many of the features present here (negative pressure shells, upper branch solutions which merge with the `hook' solutions, paperclip shapes in the strongly charged limit, etc.) are independent of dyonic charge distribution model and rather are general features of dyonically charged quark stars in pure QTEM.


\section{Stability}\label{sec:stability}

To thoroughly study radial stability of a fluid sphere, one generally considers the displacement of an individual fluid element from its equilibrium position $r$ to a new position $r+\delta r$.   Assuming that this perturbation has  harmonic time dependence, an equation for the infinitesimal radial oscillations follows. The stability  of the system can then be studied by extracting a series of eigenfrequencies $\omega_n^2$ that solve the oscillation equation. A solution is said to be radially stable if the fundamental mode eigenfrequency $\omega_0^2$ is non-negative ($\omega_0^2\geq 0$).

For an uncharged fluid sphere governed  by general relativity, the condition $\omega_0^2 \geq 0$ is satisfied at any part of the mass-radius solution curve for which $\partial M /\partial \rho_c \geq 0$. A consequence of this is that the transition from stability to instability always occurs at the maximum mass solution for a given EOS  \cite{glendenning_1997_compact,brillante2014,arbanil2015,zhang_2021_stellar}. However, when the fluid has a non-zero net charge this coincidence has been shown to be absent - the transition is offset either slightly before or slightly after the maximum mass point \cite{arbanil2015,goncalves2020,zhang_2021_stellar}. With this, we expect that the coincidence is also broken for QTEM stars with a dyonic charge. Even less is known about the stability properties of our exotic stars with negative pressure shells. While we leave a thorough analysis of the fundamental radial oscillation modes for future work, we briefly consider the sound speed and adiabatic index of the quark matter coupled to (dark) QTEM.  

The causality condition ensures that in the interior of a stable star the speed of sound $c_s=\sqrt{\left(\frac{\partial p}{\partial \rho}\right)_S}$ never exceeds the speed of light $c$. The non-interacting quark matter equation of state \eqref{eq:EOS_unitless} employed here has a constant subluminal sound speed of $c_s=\frac{c}{\sqrt{3}}$, and thus the causality condition is always satisfied. Similarly, the adiabatic index
\begin{equation}
    \gamma_\mathrm{eff}=\left(1+\frac{\rho}{p} \right)\left(\frac{dP}{d\rho}\right)_S
\end{equation}
is often used as another indicator of stability, and a bridge between the ``relativistic structure of a spherical static object and the equation of state of the interior fluid" \cite{moustakidis2017_stability}, with the subscript $S$ indicating constant specific entropy. In principle a critical $\langle\gamma_\mathrm{eff}\rangle$ exists below which configurations are unstable against radial perturbations. In general relativity the critical value is known ($\gamma_\mathrm{cr}=\frac{4}{3}+\frac{19}{42} \beta_c$ \cite{moustakidis2017_stability,chandrasekhar1964}), where $\beta_c=2M/R=R_S/R$ is the compactness parameter. When $\beta_c\to 0$ the well-known classical Newtonian limit is recovered as expected ($\langle\gamma_\mathrm{eff}\rangle\geq\frac{4}{3}$).

There is no equivalent bound that takes into account QTEM terms. Despite this, it is common practice in the compact star literature to plot the adiabatic index of a star relative to the Newtonian critical value \cite{banerjee_2021_quark,banerjee_2021_strange,hansraj_2020_isotropic,singh_2022_anisotropic}. Since  we have  restricted our investigation   to a simple, non-interacting quark matter equation of state \eqref{eq:EOS_unitless} it is straightforward to show that $\gamma_\mathrm{eff}=\frac{1}{3}(4+\frac{1}{p(r)})$. This means that the negative pressure ($p<0$) regions of these exotic star solutions actually violate the classical Newtonian bound of $\gamma_\mathrm{eff}=\frac{4}{3}$. It is not clear whether this indicates true instability or is exposing the naivety of applying the Newtonian bound to solutions to a non-linear theory. Future work should be done to clarify this point, beginning with a full investigation of the system's fundamental mode eigenfrequencies under radial perturbation.

\section{Discussion and Summary}\label{sec:summary}
In this paper we have studied non-interacting quark star solutions in pure quasi-topological electromagnetism, which we regard as a form of dark energy. Since the electric/magnetic cases on their own lead to a vanishing QTEM term, only the dyonic case has a non-trivial contribution from the non-linear theory (as is the case with QTEM black holes \cite{Cisterna_2020,Barrientos_2022,Li_2022}). With this, the effect of adding a small dyonic charge density (under either charge model) to the system is the inflation of the `normal' set of $M/R$ curves to larger masses/radii while also inducing a second branch at large mass and radius representing the solutions which include the outer shell of negative pressure. As the charge parameter becomes relatively large these two branches merge in $M/R$ space and leave a gap where a certain range of central densities have no quark star solutions. When the charge parameter becomes sufficiently large (and many of the central densities aren't associated with solutions), we finally see a characteristic `paperclip' shape   appearing in $M/R$ space, regardless of charge model.

If the  stars described in this paper were to exist in our universe, they would be electromagnetically invisible to us since there is no coupling to Maxwell's field, and as such, could only be detected   gravitationally.  
An obvious extension of our work would
be to consider dark dyonic quark stars that take into account quark matter interactions \cite{zhang_2021_unified,zhang_2021_stellar,gammon_2024}, as well as the effects of dark QTEM on  deep inelastic scattering, top quark production, and other strong interactions in particle physics.

More future work should be concerned with obtaining solutions with both $\alpha_1$ and $\alpha_2$  non-zero; this  would require a new metric ansatz.  There
should also be an investigation  of the stability of these  quark stars under radial perturbations.  While it seems   likely that the upper branch solutions are unstable, a proper analysis needs to be carried out.  
Similarly, it would be interesting to see how Buchdahl's bound \cite{buchdahl_1959} is modified in the presence of the non-linear quasi-topological terms. It is known to be modified by a pure electric charge in Maxwell electromagnetism \cite{bohmer_2007}, and is also modified in interesting ways in higher curvature gravity theories \cite{chakraborty_2020,gammon_2024,gammon_2025_chargedquarkstars4degb}. A similar bound that takes into account  the presence of QTEM terms could help shine some light on whether the upper branch   solutions close to the black hole horizon should be considered stable physical solutions.  

A more venturesome line of research would involve an exploration of the possibility that these dark dyonic QTEM quark stars could form a substantial component of the dark matter and dark energy of the universe.  The exterior metric \eqref{vacanz} is not asymptotically flat -- each star generates an effective cosmological constant given by the last term in
\eqref{vacanz}.  It is conceivable that the collective effect of such objects is to contribute to the overall dark energy of the universe.   Whether such a speculative idea is viable is a subject for future study.

\section*{Acknowledgements}
This work was supported in part by the Natural Sciences and Engineering Research Council of Canada.

\bibliographystyle{unsrt}

\cleardoublepage
\phantomsection  
\renewcommand*{\refname}{References}

\bibliography{refs}

\end{document}